\documentstyle[aps,axodraw,psfig,float]{revtex}

\newcommand{\ba}{\begin{eqnarray}}
\newcommand{\ea}{\end{eqnarray}}

\newcommand{\ep}{\epsilon}
\newcommand{\Li}{{\rm Li}}

\newcommand{\be}{\begin{equation}}
\newcommand{\ee}{\end{equation}}

\def\MSbar{$\overline{{\rm MS}}$}

\def\O#1{{\cal O}\left(#1\right)}
\def\M{m_H^2}

\def\measure{\int \frac{d^dk}{(2\pi)^{d}}\frac{d^dl}{(2\pi)^{d}} }
\def\measureVV{\int
  \frac{d^dk}{i\pi^{\frac{d}{2}}}\frac{d^dl}{i\pi^{\frac{d}{2}}} }
\def\measureVR{\int
  \frac{d^dk}{i\pi^{\frac{d}{2}}} d^dl }
\def\measureRR{\int d^dk d^dl }
\def\LOOP{{\cal L}}
\def\PS{{\cal P}}
\def\Sab#1{{\rm S}_{12}\left(#1\right)}
\def\lit#1{{\rm Li}_{3}\left(#1\right)}
\def\lid#1{{\rm Li}_{2}\left(#1\right)}
\def\zt{\zeta_3}
\def\zd{\zeta_2}
\def\zq{\zeta_4}

\def\xtri#1{
\begin{picture}(80, 60)(0, #1)
\Text(5,10)[]{$p_2$}
\Text(5,50)[]{$p_1$}
\ArrowLine(10,10)(20, 10)
\ArrowLine(10,50)(20,50)
\Line(20,10)(60,30)
\Line(20,50)(60,30)
\Line(20,50)(40,20)
\Line(20,10)(40,40)
\DashLine(70,40)(70,20){4}
\SetWidth{2}
\Line(60,30)(80,30)
\end{picture}
}

\def\tri#1{
\begin{picture}(80, 60)(0, #1)
\Text(5,10)[]{$p_2$}
\Text(5,50)[]{$p_1$}
\ArrowLine(10,10)(20, 10)
\ArrowLine(10,50)(20,50)
\Line(20,10)(60,30)
\Line(20,50)(60,30)
\DashLine(70,40)(70,20){4}
\Oval(20,30)(20,8)(0)
\SetWidth{2}
\Line(60,30)(80,30)
\end{picture}
}

\def\suns#1{
\begin{picture}(80, 60)(0, #1)
\Text(5,20)[]{$p_2$}
\Text(5,40)[]{$p_1$}
\ArrowLine(10,20)(20, 30)
\ArrowLine(10,40)(20,30)
\Line(20,30)(60,30)
\CArc(40,30)(20,0,360)
\DashLine(67,40)(67,20){4}
\SetWidth{2}
\Line(60,30)(74,30)
\end{picture}
}

\def\sunsb#1{
\begin{picture}(80, 60)(0, #1)
\Line(10,20)(20, 30)
\Line(10,40)(20,30)
\Line(20,30)(60,30)
\CArc(40,30)(20,0,360)
\DashLine(67,40)(67,20){4}
\Line(74,30)(80, 40)
\Line(74,30)(80, 20)
\SetWidth{2}
\Line(60,30)(74,30)
\end{picture}
}

\def\glass#1{
\begin{picture}(80, 60)(0, #1)
\Text(5,20)[]{$p_2$}
\Text(5,40)[]{$p_1$}
\ArrowLine(10,20)(20, 30)
\ArrowLine(10,40)(20,30)
\CArc(30,30)(10,0,360)
\CArc(50,30)(10,0,360)
\DashLine(70,40)(70,20){4}
\SetWidth{2}
\Line(60,30)(80,30)
\end{picture}
}

\def\glassb#1{
\begin{picture}(80, 60)(0, #1)
\Line(10,20)(20, 30)
\Line(10,40)(20,30)
\CArc(30,30)(10,0,360)
\CArc(50,30)(10,0,360)
\DashLine(67,40)(67,20){4}
\Line(74,30)(80, 40)
\Line(74,30)(80, 20)
\SetWidth{2}
\Line(60,30)(74,30)
\end{picture}
}

\def\cglass#1{
\begin{picture}(90, 60)(0, #1)
\Text(5,20)[]{$p_2$}
\Text(5,40)[]{$p_1$}
\Text(87,20)[]{$p_2$}
\Text(87,40)[]{$p_1$}
\ArrowLine(70,30)(80, 40)
\ArrowLine(70,30)(80,20)
\ArrowLine(10,20)(20, 30)
\ArrowLine(10,40)(20,30)
\CArc(30,30)(10,0,360)
\CArc(60,30)(10,0,360)
\DashLine(45,40)(45,20){4}
\SetWidth{2}
\Line(40,30)(50,30)
\end{picture}
}

\def\YA#1{
\begin{picture}(80,60)(0,#1)
\Text(5,20)[]{$p_2$}
\Text(5,40)[]{$p_1$}
\Text(73,10)[]{$p_2$}
\Text(73,50)[]{$p_1$}
\ArrowLine(10,20)(20, 30)
\ArrowLine(10,40)(20,30)
\ArrowLine(55,50)(65,50)
\ArrowLine(55,10)(65,10)
\Line(20,30)(55,10)
\Oval(55,30)(20,5)(0)
\DashLine(37.5,50)(37.5,10){4}
\SetWidth{2}
\Line(20, 30)(55, 50)
\end{picture}
}

\def\YB#1{
\begin{picture}(80,60)(0,#1)
\Text(5,20)[]{$p_2$}
\Text(5,40)[]{$p_1$}
\Text(73,20)[]{$p_2$}
\Text(73,40)[]{$p_1$}
\ArrowLine(10,20)(20, 30)
\ArrowLine(10,40)(20,30)
\ArrowLine(55,30)(65,40)
\ArrowLine(55,30)(65,20)
\CArc(37.5, 30)(17.5, 180, 450)
\CArc(55,47.5)(17.5,180,270)
\DashLine(32.5,50)(32.5,10){4}
\SetWidth{2}
\CArc(37.5, 30)(17.5, 90, 180)
\end{picture}
}

\def\YC#1{
\begin{picture}(80,60)(0,#1)
\Text(5,20)[]{$p_2$}
\Text(5,40)[]{$p_1$}
\Text(73,10)[]{$p_2$}
\Text(73,50)[]{$p_1$}
\ArrowLine(10,20)(20, 30)
\ArrowLine(10,40)(20,30)
\ArrowLine(55,50)(65,50)
\ArrowLine(55,10)(65,10)
\Line(20,30)(55,10)
\Line(45,44.28)(55,50)
\Line(45,44.28)(55,10)
\Line(55,50)(55,10)
\DashLine(37.5,50)(37.5,10){4}
\SetWidth{2}
\Line(20, 30)(45, 44.28)
\end{picture}
}

\def\YE#1{
\begin{picture}(80, 60)(0, #1)
\Text(5,20)[]{$p_2$}
\Text(5,40)[]{$p_1$}
\Text(75,20)[]{$p_2$}
\Text(75,40)[]{$p_1$}
\ArrowLine(10,20)(20, 30)
\ArrowLine(10,40)(20,30)
\ArrowLine(60,30)(70,40)
\ArrowLine(60,30)(70,20)
\CArc(30,30)(10,180,360)
\CArc(50,30)(10,0,360)
\DashLine(30,50)(30,10){4}
\SetWidth{2}
\CArc(30,30)(10,0,180)
\end{picture}
}

\def\YD#1{
\begin{picture}(80,60)(0,#1)
\Text(5,10)[]{$p_2$}
\Text(5,50)[]{$p_1$}
\Text(73,10)[]{$p_1$}
\Text(73,50)[]{$p_2$}
\ArrowLine(10,50)(20, 50)
\ArrowLine(10,10)(20,10)
\ArrowLine(55,50)(65,50)
\ArrowLine(55,10)(65,10)
\Line(20,10)(55,10)
\Line(55,50)(55,10)
\Line(37.5,50)(37.5,10)
\Line(20,50)(20,10)
\Line(35, 50)(55, 50)
\DashLine(28.25,55)(28.25,5){4}
\SetWidth{2}
\Line(20, 50)(37.5, 50)
\end{picture}
}

\def\YF#1{
\begin{picture}(80,60)(0,#1)
\Text(5,20)[]{$p_2$}
\Text(5,40)[]{$p_1$}
\Text(73,10)[]{$p_2$}
\Text(73,50)[]{$p_1$}
\ArrowLine(10,20)(20, 30)
\ArrowLine(10,40)(20,30)
\ArrowLine(55,50)(65,50)
\ArrowLine(55,10)(65,10)
\Line(20,30)(55,10)
\Line(45,44.28)(55,50)
\Line(45,44.28)(55,10)
\Line(55,50)(45,16.72)
\DashLine(37.5,50)(37.5,10){4}
\SetWidth{2}
\Line(20, 30)(45, 44.28)
\end{picture}
}

\def\YAb#1{
\begin{picture}(80,60)(0,#1)
\Line(10,20)(20, 30)
\Line(10,40)(20,30)
\Line(55,50)(65,50)
\Line(55,10)(65,10)
\Line(20,30)(55,10)
\Oval(55,30)(20,5)(0)
\DashLine(37.5,50)(37.5,10){4}
\SetWidth{2}
\Line(20, 30)(55, 50)
\end{picture}
}

\def\YBb#1{
\begin{picture}(80,60)(0,#1)
\Line(10,20)(20, 30)
\Line(10,40)(20,30)
\Line(55,30)(65,40)
\Line(55,30)(65,20)
\CArc(37.5, 30)(17.5, 180, 450)
\CArc(55,47.5)(17.5,180,270)
\DashLine(32.5,50)(32.5,10){4}
\SetWidth{2}
\CArc(37.5, 30)(17.5, 90, 180)
\end{picture}
}

\def\YCb#1{
\begin{picture}(80,60)(0,#1)
\Line(10,20)(20, 30)
\Line(10,40)(20,30)
\Line(55,50)(65,50)
\Line(55,10)(65,10)
\Line(20,30)(55,10)
\Line(45,44.28)(55,50)
\Line(45,44.28)(55,10)
\Line(55,50)(55,10)
\DashLine(37.5,50)(37.5,10){4}
\SetWidth{2}
\Line(20, 30)(45, 44.28)
\end{picture}
}

\def\YEb#1{
\begin{picture}(80, 60)(0, #1)
\Line(10,20)(20, 30)
\Line(10,40)(20,30)
\Line(60,30)(70,40)
\Line(60,30)(70,20)
\CArc(30,30)(10,180,360)
\CArc(50,30)(10,0,360)
\DashLine(30,50)(30,10){4}
\SetWidth{2}
\CArc(30,30)(10,0,180)
\end{picture}
}

\def\Xone#1{
\begin{picture}(80,60)(0,#1)
\Text(5,20)[]{$p_2$}
\Text(5,40)[]{$p_1$}
\Text(77,20)[]{$p_2$}
\Text(77,40)[]{$p_1$}
\Line(10,20)(20, 30)
\Line(10,40)(20,30)
\Line(60,30)(70,40)
\Line(60,30)(70,20)
\CArc(40,30)(20,0,360)
\DashLine(40,55)(40,5){4}
\SetWidth{2}
\Line(20, 30)(60, 30)
\end{picture}
}

\def\Xeleven#1{
\begin{picture}(80,60)(0,#1)
\Text(5,10)[]{$p_2$}
\Text(5,50)[]{$p_1$}
\Text(73,10)[]{$p_2$}
\Text(73,50)[]{$p_1$}
\ArrowLine(10,50)(20, 50)
\ArrowLine(10,10)(20,10)
\ArrowLine(55,50)(65,50)
\ArrowLine(55,10)(65,10)
\Line(20,10)(55,10)
\Line(55,50)(55,10)
\Line(37.5,50)(37.5,10)
\Line(20,50)(20,10)
\Line(35, 50)(55, 50)
\DashLine(26,55)(49,5){4}
\Line(20,10)(55,10)
\SetWidth{2}
\Line(20, 50)(37.5, 50)
\end{picture}
}

\def\Xfourteen#1{
\begin{picture}(80,60)(0,#1)
\Text(5,20)[]{$p_2$}
\Text(5,40)[]{$p_1$}
\Text(73,20)[]{$p_2$}
\Text(73,40)[]{$p_1$}
\Line(10,20)(20, 30)
\Line(10,40)(20,30)
\Line(55,30)(65,40)
\Line(55,30)(65,20)
\DashCArc(37.5, 30)(17.5, 90, 180){1.5}
\CArc(37.5, 30)(17.5, 180, 450)
\DashLine(44.5,50)(44.5,10){4}
\SetWidth{2}
\CArc(55,47.5)(17.5,180,270)
\end{picture}
}

\def\Xten#1{
\begin{picture}(80,60)(0,#1)
\Text(5,20)[]{$p_2$}
\Text(5,40)[]{$p_1$}
\Text(73,20)[]{$p_2$}
\Text(73,40)[]{$p_1$}
\Line(10,20)(20, 30)
\Line(10,40)(20,30)
\Line(55,30)(65,40)
\Line(55,30)(65,20)
\CArc(37.5, 30)(17.5, 0, 360)
\DashLine(30,50)(45,10){4}
\SetWidth{2}
\Line(37.5, 47.50)(37.5,12.50)
\end{picture}
}

\def\Xthirteen#1{
\begin{picture}(80,60)(0,#1)
\Text(5,10)[]{$p_2$}
\Text(5,50)[]{$p_1$}
\Text(73,10)[]{$p_1$}
\Text(73,50)[]{$p_2$}
\ArrowLine(10,50)(20, 50)
\ArrowLine(10,10)(20,10)
\ArrowLine(55,50)(65,50)
\ArrowLine(55,10)(65,10)
\Line(20,10)(55,10)
\Line(55,50)(55,10)
\Line(20,50)(20,10)
\Line(35, 50)(55, 50)
\DashLine(26,55)(49,5){4}
\Line(20, 50)(37.5, 50)
\SetWidth{2}
\Line(37.5,50)(37.5,10)
\end{picture}
}

\def\Xtwentyfive#1{
\begin{picture}(80,60)(0,#1)
\Text(5,10)[]{$p_2$}
\Text(5,50)[]{$p_1$}
\Text(73,10)[]{$p_1$}
\Text(73,50)[]{$p_2$}
\ArrowLine(10,50)(20, 50)
\ArrowLine(10,10)(20,10)
\ArrowLine(55,50)(65,50)
\ArrowLine(55,10)(65,10)
\Line(20,10)(55,10)
\Line(55,50)(55,10)
\Line(37.5,50)(37.5,10)
\Line(20,50)(20,10)
\Line(35, 50)(55, 50)
\DashLine(26,55)(49,5){4}
\Line(20,10)(55,10)
\SetWidth{2}
\Line(20, 50)(37.5, 50)
\end{picture}
}
\def\Xfive#1{
\begin{picture}(80,60)(0,#1)
\Text(5,10)[]{$p_2$}
\Text(5,50)[]{$p_1$}
\Text(73,10)[]{$p_1$}
\Text(73,50)[]{$p_2$}
\ArrowLine(10,50)(20, 50)
\ArrowLine(10,10)(20,10)
\ArrowLine(55,50)(65,50)
\ArrowLine(55,10)(65,10)
\Line(20,10)(55,10)
\Line(55,50)(55,10)
\Line(20,30)(55,30)
\Line(20,50)(20,10)
\DashLine(37.5,55)(37.5,5){4}
\Line(20,10)(55,10)
\SetWidth{2}
\Line(20, 50)(55, 50)
\end{picture}
}
\def\Xeight#1{
\begin{picture}(80,60)(0,#1)
\Text(5,10)[]{$p_2$}
\Text(5,50)[]{$p_1$}
\Text(73,10)[]{$p_1$}
\Text(73,50)[]{$p_2$}
\ArrowLine(10,50)(20, 50)
\ArrowLine(10,10)(20,10)
\ArrowLine(55,50)(65,50)
\ArrowLine(55,10)(65,10)
\Line(20,10)(55,10)
\Line(55,50)(55,10)
\Line(20,50)(20,10)
\DashLine(37.5,55)(37.5,5){4}
\Line(20,10)(55,10)
\Line(20, 50)(55, 50)
\SetWidth{2}
\Line(20,30)(55,30)
\end{picture}
}

\def\Xthree#1{
\begin{picture}(80,60)(0,#1)
\Text(5,10)[]{$p_2$}
\Text(5,50)[]{$p_1$}
\Text(73,10)[]{$p_1$}
\Text(73,50)[]{$p_2$}
\ArrowLine(10,50)(20, 50)
\ArrowLine(10,10)(20,10)
\ArrowLine(55,50)(65,50)
\ArrowLine(55,10)(65,10)
\Line(55,50)(55,10)
\Line(20,50)(20,10)
\DashLine(37.5,55)(37.5,0){3.5}
\Oval(37.5,10)(4,17.5)(0)
\SetWidth{2}
\Line(20, 50)(55, 50)
\end{picture}
}

\def\Xsix#1{
\begin{picture}(80,60)(0,#1)
\Text(5,10)[]{$p_2$}
\Text(5,50)[]{$p_1$}
\Text(73,10)[]{$p_1$}
\Text(73,50)[]{$p_2$}
\ArrowLine(10,50)(20, 50)
\ArrowLine(10,10)(20,10)
\ArrowLine(55,50)(65,50)
\ArrowLine(55,10)(65,10)
\Line(55,50)(55,10)
\Line(20,50)(20,10)
\DashLine(40,55)(40,0){3.5}
\Line(20, 28)(55, 10)
\Line(20, 10)(55, 32)
\SetWidth{2}
\Line(20, 50)(55, 50)
\end{picture}
}

\def\Xthirty#1{
\begin{picture}(80,60)(0,#1)
\Text(5,20)[]{$p_2$}
\Text(5,40)[]{$p_1$}
\Text(73,10)[]{$p_2$}
\Text(73,50)[]{$p_1$}
\Line(10,20)(20, 30)
\Line(10,40)(20,30)
\Line(55,50)(65,50)
\Line(55,10)(65,10)
\Line(20,30)(55,10)
\Line(45,44.28)(55,50)
\Line(45,44.28)(55,10)
\Line(55,50)(45,16.72)
\DashLine(37,55)(52,5){4}
\SetWidth{2}
\Line(20, 30)(45, 44.28)
\end{picture}
}

\def\Xtwentyone#1{
\begin{picture}(80,60)(0,#1)
\Text(5,10)[]{$p_2$}
\Text(5,50)[]{$p_1$}
\Text(73,10)[]{$p_1$}
\Text(73,50)[]{$p_2$}
\ArrowLine(10,50)(20, 50)
\ArrowLine(10,10)(20,10)
\ArrowLine(55,50)(65,50)
\ArrowLine(55,10)(65,10)
\Line(55,50)(55,10)
\Line(20,50)(20,10)
\DashLine(40,55)(40,0){3.5}
\Line(20, 28)(55, 10)
\Line(20, 50)(55, 50)
\SetWidth{2}
\Line(20, 10)(55, 32)
\end{picture}
}

\def\Xeighteen#1{
\begin{picture}(80,60)(0,#1)
\Text(5,20)[]{$p_2$}
\Text(5,40)[]{$p_1$}
\Text(73,10)[]{$p_2$}
\Text(73,50)[]{$p_1$}
\ArrowLine(10,20)(20, 30)
\ArrowLine(10,40)(20,30)
\ArrowLine(55,50)(65,50)
\ArrowLine(55,10)(65,10)
\Line(20,30)(55,10)
\Line(45,44.28)(55,50)
\Line(45,44.28)(55,10)
\Line(20, 30)(45, 44.28)
\DashLine(37,55)(52,5){4}
\SetWidth{2}
\Line(55,50)(45,16.72)
\end{picture}
}

\def\Xtwenty#1{
\begin{picture}(80,60)(0,#1)
\Text(5,10)[]{$p_2$}
\Text(5,50)[]{$p_1$}
\Text(73,10)[]{$p_2$}
\Text(73,50)[]{$p_1$}
\ArrowLine(10,10)(20, 10)
\ArrowLine(10,50)(20,50)
\ArrowLine(55,50)(65,50)
\ArrowLine(55,10)(65,10)
\Line(20,10)(55,10)
\Line(40,50)(55,10)
\Line(20, 50)(55, 50)
\Line(20, 50)(20,10)
\DashLine(32,55)(51,5){4}
\SetWidth{2}
\Line(55,50)(40,10)
\end{picture}
}

\def\Xnine#1{
\begin{picture}(80,60)(0,#1)
\Text(5,20)[]{$p_2$}
\Text(5,40)[]{$p_1$}
\Text(73,10)[]{$p_2$}
\Text(73,50)[]{$p_1$}
\ArrowLine(10,20)(20, 30)
\ArrowLine(10,40)(20,30)
\ArrowLine(55,50)(65,50)
\ArrowLine(55,10)(65,10)
\Line(20,30)(55,10)
\Line(45,44.28)(55,50)
\Line(20, 30)(45, 44.28)
\Line(55,50)(55,10)
\DashLine(37,55)(52,5){4}
\SetWidth{2}
\Line(55,50)(37.5,20)
\end{picture}
}

\def\Xseven#1{
\begin{picture}(80,60)(0,#1)
\Text(5,20)[]{$p_2$}
\Text(5,40)[]{$p_1$}
\Text(73,10)[]{$p_2$}
\Text(73,50)[]{$p_1$}
\ArrowLine(10,20)(20, 30)
\ArrowLine(10,40)(20,30)
\ArrowLine(55,50)(65,50)
\ArrowLine(55,10)(65,10)
\Line(20,30)(55,10)
\Line(45,44.28)(55,50)
\Line(20, 30)(45, 44.28)
\Line(55,50)(55,10)
\DashLine(37,55)(52,5){4}
\SetWidth{2}
\Line(20,30)(55,30)
\end{picture}
}

\def\Xsixteen#1{
\begin{picture}(80,60)(0,#1)
\Text(5,20)[]{$p_2$}
\Text(5,40)[]{$p_1$}
\Text(73,10)[]{$p_2$}
\Text(73,50)[]{$p_1$}
\ArrowLine(10,20)(20, 30)
\ArrowLine(10,40)(20,30)
\ArrowLine(55,50)(65,50)
\ArrowLine(55,10)(65,10)
\Line(20,30)(55,50)
\Line(20,30)(37.5,20)
\Line(20, 30)(45, 44.28)
\Line(55,50)(55,10)
\Line(55,50)(37.5,20)
\DashLine(37,55)(52,5){4}
\SetWidth{2}
\Line(37.5, 20)(55,10)
\end{picture}
}

\def\Xnineteen#1{
\begin{picture}(80,60)(0,#1)
\Text(5,10)[]{$p_2$}
\Text(5,50)[]{$p_1$}
\Text(73,10)[]{$p_2$}
\Text(73,50)[]{$p_1$}
\ArrowLine(10,10)(20, 10)
\ArrowLine(10,50)(20,50)
\ArrowLine(55,50)(65,50)
\ArrowLine(55,10)(65,10)
\Line(20,10)(55,10)
\Line(55,50)(55,10)
\Line(20, 50)(55, 50)
\DashLine(20, 50)(20,10){2}
\DashLine(32,55)(51,5){4}
\SetWidth{2}
\Line(55,50)(40,10)
\end{picture}
}

\def\slash#1{\setbox0=\hbox{$#1$}               
   \dimen0=\wd0                                 
   \setbox1=\hbox{/} \dimen1=\wd1               
   \ifdim\dimen0>\dimen1                        
      \rlap{\hbox to \dimen0{\hfil/\hfil}}      
      #1                                        
   \else                                        
      \rlap{\hbox to \dimen1{\hfil$#1$\hfil}}   
      /                                         
   \fi}                                         %

\begin{document}

\title{
\[ \vspace{-2cm} \]
\noindent\hfill\hbox{\rm  SLAC-PUB-9273} \vskip 1pt
\noindent\hfill\hbox{\rm hep-ph/0207004} \vskip 10pt
Higgs boson production at hadron colliders in NNLO QCD
 }

\author{ Charalampos Anastasiou\thanks{email: babis@slac.stanford.edu} 
and Kirill Melnikov\thanks{
e-mail:  melnikov@slac.stanford.edu}}
\address{Stanford Linear Accelerator Center\\
Stanford University, Stanford, CA 94309}
\maketitle

\begin{abstract}
We compute the total cross-section for direct Higgs boson production
in hadron collisions at NNLO in perturbative  QCD.  
A new technique which allows us to perform an  algorithmic evaluation 
of inclusive phase-space integrals is introduced, based on the Cutkosky rules, 
integration by parts and the differential equation method 
for computing master integrals. Finally, we 
discuss the  numerical impact of the  ${\cal O}(\alpha_s^2)$ QCD corrections 
to the Higgs boson production cross-section at the LHC and the Tevatron.
\end{abstract}


\section{Introduction}
\label{sec:intro}
The Higgs boson is currently the only missing particle in the minimal 
Standard Model (SM) of  electroweak interactions. Its discovery 
will be one of the final steps toward the  
experimental verification of the SM, 
and will  provide  useful input for detailed studies of 
the mass generation  mechanism and for physics beyond the SM.  

Direct searches at LEP restrict the Higgs boson mass to be greater 
than $114.1~{\rm GeV}$~\cite{higgslep}, while  a global fit to precision 
electroweak measurements~\cite{SMfit} favors a value around $90~{\rm GeV}$. 
In addition, the requirement  that the SM remains  
perturbative up to relatively high energy scales sets an  upper bound 
at approximately  
$1~{\rm TeV}$~\cite{kniehl}. Although the above evidence is not completely 
conclusive, it indicates a relatively light Higgs boson  which could be 
observed  at either the Tevatron  or the LHC. At both of these facilities,
gluon fusion through top-quark loops is expected to be the dominant Higgs 
production mechanism. All other channels, such as  vector boson fusion   
$q q \to H qq$ and associated Higgs production $q  \bar q \prime \to HW$, 
are suppressed by about an order of magnitude 
(see Ref.\cite{spira1} for a review). We therefore focus upon 
the process $gg \to H$ in this paper.

The theoretical estimates of the cross-section for the Higgs boson production via gluon fusion, 
based on computations through to the next-to-leading order (NLO) in perturbative QCD,
turn out to be insufficient. The leading-order (LO) cross-section 
is proportional to $\alpha_s^2(\mu^2)$, and for this reason exhibits a 
strong dependence on the choice of the scale $\mu$. Including the 
${\cal O}(\alpha_s)$ corrections~\cite{oneloopd,oneloop} decreases the scale 
dependence, but the cross-section increases by  a very large amount, approximately $70\%$.   
It is therefore important  to evaluate  the next order 
in the perturbative expansion, since this is the only way
to enhance the credibility of the theoretical predictions. 

To compute the cross-section to next-to-next-to-leading  
order (NNLO), we must combine:  
the matrix elements  for the ${\cal O}(\alpha_s^2)$  
virtual corrections  to $gg \to H$; the matrix elements for the 
${\cal O}(\alpha_s)$ virtual  corrections  to $gg \to H g$,  $q g \to H q$, 
and $q \bar q \to H g$; and finally the tree-level matrix elements for
the  processes $gg \to Hgg$, $ gg \to H q \bar q$, $qg \to H q g$, 
$q \bar q \to H g g$, and  $q \bar q \to H q \bar q$. 
For the inclusive cross-section we must integrate over the 
loop-momenta in the virtual amplitudes and the phase-space of the real 
particles in the final state. Both real and virtual corrections are 
divergent in four dimensions. We regularize the amplitudes using 
conventional dimensional regularization ($d=4-2\epsilon$), and remove 
the ultraviolet divergences by renormalizing 
in the \MSbar\ scheme. The remaining divergences
arise from initial state collinear radiation and are absorbed into the 
parton distribution functions, yielding a finite cross-section.  

The calculation can be simplified substantially by considering the 
limit where the Higgs boson is much lighter than twice the 
mass of the top-quark. In this limit, the top-quark loops are 
replaced by point-like  vertices. The corresponding effective Lagrangian 
is known to provide a satisfactory description of the cross-section for a 
light Higgs boson at NLO~\cite{oneloopd,oneloop,refadd}. 

In the heavy top-quark limit, the NNLO contributions to
the direct Higgs production cross-section are topologically similar to the 
${\cal O}(\alpha_s^2)$ corrections to the  Drell-Yan process
which have been calculated in the past~\cite{drellyan}. The 
phase-space and loop integrals required for the calculation of 
the Higgs boson production cross-section  could in principle be obtained 
in a similar fashion. However, such an approach is impractical for the Higgs
production cross-section due to the larger number of Feynman diagrams with 
a considerably more complicated tensor structure. 
For a problem of this complexity a highly automated algorithm which treats
virtual and real corrections in a unified manner is  desirable. 
 
It is  well known how to construct algorithms which in principle 
can perform multi-loop integrations. First, one can employ the 
method of integration by parts (IBP)~\cite{IBP} 
in order to reduce the number of integrals involved in such computations. 
Algorithms which find the solutions of IBP identities 
in a process and topology independent manner are available~\cite{LI,laporta}. 
After the application of IBP, a small number of remaining integrals 
which are not reducible further (master integrals), must be evaluated 
explicitly. Powerful techniques such as the differential
equation method~\cite{kotik,LI} and the Mellin-Barnes integral 
representation~\cite{mellinbarnes} can then be employed
to derive an expansion of the master integrals in $\epsilon$.  
The above methods provide  general purpose tools 
for the evaluation of virtual corrections. However, similar  
methods do not exist for computing phase-space integrals; they  
are usually calculated manually and on a case by case 
basis. In this paper we present 
an algorithmic procedure
for evaluating phase-space integrals, based on the 
Cutkosky rules~\cite{cutkosky}, integration  by parts~\cite{IBP} and
the differential equation method~\cite{LI}.

Partial results for the NNLO corrections to the Higgs boson production 
cross-section are available in the literature. 
The NNLO virtual corrections were  computed in~\cite{harlander} 
by Harlander. The ``soft''  part of the cross-section at NNLO was derived  in~\cite{hksoft,catani} by  extracting  
contributions to the partonic cross-sections that  are  singular
when the partonic center of mass energy $\sqrt{\hat {s}}$
equals the mass of the Higgs boson $m_H$. Recently,  
Harlander and Kilgore~\cite{hkhard}  obtained an excellent 
approximation to the complete NNLO result
by expanding  the phase-space integrals around the 
kinematic point $\hat s = \M$.
In this paper we present  the full analytic result for the NNLO
corrections to the Higgs boson production cross-section. In our derivation 
we do not need to resort to an expansion around a special kinematic point
and our expressions are therefore valid for an arbitrary ratio $m_H^2/\hat s$.

The paper is organized as follows. In Section~\ref{sec:notation} we  
briefly  review the effective Lagrangian for describing gluon 
interactions with  the Higgs boson. We also introduce our notations 
and present 
all  basic formulae and definitions for the total cross-section.
In Section~\ref{sec:method} we describe our method for solving 
multi-particle phase-space integrals in an algorithmic fashion and illustrate
its application with a few typical examples. We present the analytic 
expressions for the renormalized partonic cross-sections   $ij \to H + X$
in Section~\ref{sec:results}. 
In Section~\ref{sec:numerics} we discuss the impact of the 
${\cal O}(\alpha_s^2)$ corrections on the Higgs boson production 
cross-section at the Tevatron and at the LHC. 
We present our conclusions  in Section~\ref{sec:conclusions}.
Some useful formulae, including the complete list of master integrals, 
are collected in the Appendix.

\section{Effective Lagrangian}
\label{sec:notation}

The Higgs boson interaction with gluons is a loop induced process and is
therefore sensitive to 
all colored particles  which get their masses through the Higgs mechanism. 
In this paper  we restrict ourselves to the Standard Model where the top-quark  contribution dominates. 

Although  the  Born cross-section is known as a function of 
the top mass $m_t$  and the Higgs boson mass  $ m_H$,  
it is much harder to obtain the 
exact analytic  dependence of the cross-section on the mass of the top-quark
in higher orders of perturbation theory.
However, since  it is most probable that the Higgs boson is light, 
it is  sufficient to work in the infinite top-quark mass limit.

For Higgs  boson masses in the range $100-200~{\rm GeV}$,
we can  describe the Higgs gluon interaction by introducing the effective 
Lagrangian~\cite{oneloopd,oneloop,refadd}
\be
{\cal L}_{\rm eff} = - \frac{1}{4v} C_1 G_{\mu \nu}^a G^{a \mu \nu} H, 
\label{eff}
\ee 
where $G_{\mu \nu}^a$ is the gluon strength tensor, $H$ is 
the Higgs field and $v \approx 246~{\rm GeV}$ 
is the Higgs boson vacuum expectation 
value. The Wilson coefficient $C_1$, defined in the 
$\overline {\rm MS}$ scheme, is~\cite{rencon} 
\be
C_1 = \frac{-1}{3\pi} \left \{ 1 + \frac{11}{4} \frac{\alpha_s}{\pi} 
+ \left (\frac{\alpha_s}{\pi}  \right)^2 \left [ 
\frac{2777}{288} + \frac{19}{16} L_t + n_f \left( - \frac{67}{96} 
+ \frac{1}{3} L_t \right ) \right ] + {\cal O}(\alpha_s^3) \right \},
\ee
where $\alpha_s(\mu)$ is the \MSbar\ 
strong coupling constant, 
$n_f = 5$ is the number of active flavors and 
$L_t = \log(\mu^2/m_t^2)$.

It is expected that  the effective Lagrangian of Eq.~(\ref{eff}) is a valid 
approximation to the Higgs gluon interaction for small values 
($m_H < 2m_t$) of the 
Higgs boson mass. It can be checked that at leading order and for  
$m_H \sim 150~{\rm GeV}$, the effective Lagrangian approximation 
is accurate within $5\%$, whereas for $m_H \sim 200~{\rm GeV}$, the accuracy 
drops to $10 \%$. The precision of the approximation improves  
for the Higgs boson production cross-section computed at 
NLO accuracy~\cite{oneloop}.
The effective Lagrangian description therefore seems accurate 
in the entire range of phenomenologically interesting 
Higgs boson masses, and  we adopt it for the calculation of the 
NNLO corrections.

The effective Lagrangian approach separates the short-distance 
($\sim m_t^{-1}$) and the long-distance  
($\sim m_H^{-1}$) scales, simplifying the calculation
of the Higgs boson production cross-section. For example, the  original 
one-loop triangle diagram of the gluon-gluon Higgs interaction vertex at LO is 
now replaced by the simple tree-level vertex derived from Eq.(\ref{eff}).  
The effective Lagrangian  approach also yields the correct  $Hggg$ and $Hgggg$ 
interaction vertices, as a consequence of gauge invariance.  In addition, 
in the limit of vanishing fermion masses, there is no direct interaction 
between massless quarks and the Higgs boson.

The partonic cross-sections for the production of the Higgs boson,
up to NNLO in perturbation theory, 
receive  the following contributions:
a) virtual corrections to $gg \to H$, up to ${\cal O}(\alpha_s^2)$;
b) virtual corrections to single real emission processes
$gg \to Hg,~qg \to H q,~\bar q g \to H \bar q,~q \bar q \to Hg$, up to 
${\cal O}(\alpha_s)$; and 
c) double real emission processes
$gg \to Hgg,~gg \to H q \bar q,~qg \to Hq g,~\bar q g \to H \bar q g,
q \bar q \to H gg,~q \bar q \to H q \bar q$ to LO. 
The effective Lagrangian of Eq.(\ref{eff}) 
and the corresponding 
matrix elements should be renormalized in the $\overline {\rm MS}$ scheme 
by a global renormalization factor~\cite{rencon}:
\be
Z(\alpha_s) = 1 - \frac{\alpha_s}{\pi} \frac{\beta_0}{\ep} 
+ \left (\frac{\alpha_s}{\pi} \right )^2 \left [ \frac{\beta_0^2}{\ep^2}
-\frac{\beta_1}{\ep} \right ]+ {\cal O}(\alpha_s^3),
\ee
where $\beta_0$ and $\beta_1$ are the two first coefficients of the 
QCD $\beta$-function:
\be
\beta_0 = \frac{11}{4} - \frac{1}{6}n_f,~~
\beta_1 = \frac{51}{8} - \frac{19}{24}n_f.
\ee
This additional renormalization, together with the standard renormalization 
of the strong coupling constant removes all ultraviolet 
divergences  in the cross-section. However,  since  we work in 
the approximation of  massless  colored partons in the initial state, 
the  total cross-section is not  finite even after   
the ultraviolet renormalization has been performed. 
The remaining singularities are 
associated with  the collinear radiation 
off the colliding  partons. 
It is well known that   
these singularities factorize and can be 
removed by renormalizing the parton distribution functions in a manner 
consistent with the DGLAP evolution equation. 

The general factorization formula for the 
cross-section of the Higgs boson production 
from the collision of two hadrons $h_1(p_1)$ and $h_2(p_2)$, 
is
\be
\sigma_{h_1+h_2 \to H + X} = \sum \limits_{ij} 
\int \limits_{0}^{1} {\rm d}x_1 {\rm d} x_2 f_i^{(h_1)} (x_1) f_j^{(h_2)}(x_2)
\sigma_{ij \to H}(m_H^2,x_1x_2s),
\label{basic}
\ee
where $f_i^{(h)}(x)$ is the standard distribution function for a parton $i$ 
in the hadron $h$, $\sigma_{ij}$ is the partonic cross-section 
for $i+j \to H +X$
and $s \equiv (p_1+p_2)^2$ is 
the square of the total center of mass energy of the hadron hadron collision.
Using dimensional analysis we can write the partonic cross-sections in 
terms of the single dimensionless variable 
$z=\M/{\hat s}$, 
\be
\sigma_{ij} \sim 1/v^2
g(m_H^2/{\hat s})
\ee
with $\hat s \equiv s x_1 x_2$. Introducing $x = m_H^2/s$,  we then 
rewrite 
Eq.~(\ref{basic}) in the form 
\be
\sigma_{h_1 + h_2 \to H + X} = x \sum \limits_{ij} \left [
f_i^{(h_1)} \otimes f_j^{(h_2)} \otimes \left (  \sigma_{ij}(z) /z \right )
\right ](x),
\ee
where the standard convolution $\otimes$ is defined as
\be
\left [f_1 \otimes f_2 \right ](x) = \int \limits_{0}^{1} 
{\rm d} x_1 {\rm d} x_2 f_1(x_1) f_2(x_2) \delta(x-x_1x_2).
\ee

The collinear singularities are factored out from the partonic cross-sections 
with the following procedure. Denoting the 
unrenormalized (in the sense of collinear
singularities) partonic cross-sections by $\sigma_{ij}$, 
the $\overline {\rm MS}$ 
renormalized partonic cross-sections $\hat \sigma_{ij}$ are
implicitly given by:
\be
[ \sigma_{ij}(x)/x] = \sum \limits_{k,l}  
[\hat \sigma_{kl}(z)/z] \otimes \Gamma_{ki} \otimes 
\Gamma_{lj}, 
\label{expl}
\ee
where the kernels $\Gamma_{ij}$ are
\be
\Gamma_{ij} = \delta_{ij} \delta(1-x) 
- \frac{\alpha_s}{\pi} \frac{P_{ij}^{(0)}(x)}{\ep}
+ \left ( \frac{\alpha_s}{\pi}  \right)^2 \left [ 
\frac{1}{2\ep^2} \left ( \left ( P_{ik}^{(0)} \otimes P_{kj}^{(0)} \right )(x)
+\beta_0 P^{(0)}_{ij}(x) \right ) - \frac{1}{2\ep} P^{(1)}_{ij}(x) \right ]
+\O{\alpha_s^3}.
\label{gam}
\ee
The standard space-like splitting functions 
$P_{ij}$ 
\cite{split0,split1}  are listed in the Appendix.
We can easily solve  Eq.(\ref{expl}) for 
$\hat \sigma_{ij}(x)/x = \hat \rho_{ij}(x)$ 
order by order in  $\alpha_s$.
It is convenient to  introduce a matrix notation and rewrite 
Eq.~(\ref{expl}) as
\be
\rho = \Gamma^{T} \otimes \hat \rho \otimes \Gamma,
\label{matrix}
\ee
where $\rho$ is the matrix of partonic cross-sections in  
flavor space  and $\Gamma$ is the matrix with components $\Gamma_{ij}(x)$ as in 
Eq.(\ref{gam}). We then write
\be
\Gamma =  \delta(1-x)  U  - \frac{\alpha_s}{\pi} \Gamma_1 
+ \left ( \frac{\alpha_s}{\pi}  \right)^2 \Gamma_2 +\O{\alpha_s^3},
\ee
where the matrix $U$ has the components $U_{ij} = \delta_{ig}\delta_{jg}$ and 
$\Gamma_{1,2}$ can be read off from Eq.~(\ref{gam}).
Inverting Eq.(\ref{matrix}) to obtain 
\be
\hat \rho = \left [ \Gamma^{T} \right ]^{(-1)} \otimes \rho 
\otimes \left [ \Gamma \right]^{-1},
\ee
and expanding $\hat \rho$ in $\alpha_s$ 
$$
\hat \rho  = \hat \rho^{(0)} + \frac{\alpha_s}{\pi}  \hat \rho^{(1)}
+ \left ( \frac{\alpha_s}{\pi}  \right)^2 \hat \rho^{(2)}, 
$$
we find
\ba
&& \hat \rho^{(0)} = \rho^{(0)},~~~~~~~~
\hat \rho^{(1)} = \rho^{(1)} + \Gamma_1^{T} \otimes \rho^{(0)} 
     + \rho^{(0)} \otimes \Gamma_1,
\\
&& \hat \rho^{(2)} = \rho^{(2)} - \Gamma_2^{T} \otimes \rho^{(0)} 
     - \rho^{(0)} \otimes \Gamma_2 
- \Gamma_1^{T} \otimes \rho^{(0)} \otimes \Gamma_1
+  \Gamma_1^{T} \otimes \hat \rho^{(1)} 
     + \hat \rho^{(1)} \otimes \Gamma_1.
\nonumber 
\ea

Having derived the finite partonic cross-sections 
$\hat \sigma_{ij}$, we must convolute them with the $\overline {\rm MS}$ 
parton distribution functions $\bar f_i$ to obtain the total hadronic
cross-section:
\be
\sigma_{h_1 + h_2 \to H + X} = x \sum \limits_{ij} \left [
\bar f_i^{(h_1)} \otimes \bar f_j^{(h_2)} \otimes 
\left ( \hat  \sigma_{ij}(z) /z \right )
\right ](x).
\label{eq14}
\ee
We  present  our results for the  partonic cross-sections 
$\hat{\sigma}_{ij}(z)$ in Section~\ref{sec:results}. 
In Section ~\ref{sec:numerics} 
we  use Eq.(\ref{eq14}) to calculate 
 the Higgs boson production cross-section at the Tevatron and at the LHC.

\section{Method}
\label{sec:method}
In this Section we  describe the method employed to compute the partonic
cross-sections to NNLO. 
At this order we must calculate  three distinct contributions:
\begin{itemize}
\item  {\bf double-virtual:} the interference of the Born and  the two-loop 
        amplitude as well as the self-interference of the one-loop amplitude 
       for $gg \rightarrow H$, 
\begin{center}
\begin{picture}(300,37)(0,0)
\Gluon(40,5)(60,5){2}{4}
\Gluon(60,5)(60,35){2}{6}
\Gluon(40,35)(60,35){2}{4}
\Gluon(60,35)(100, 20){2}{8}
\Gluon(60,5)(100, 20){2}{8}
\Gluon(80,27.5)(80,12.5){2}{2}
\put(97,17){{$\otimes$}}
\SetWidth{1}
\Line(106,20)(115,20)
\Line(130,20)(137,20)
\SetWidth{0.5}
\put(135,17){{ $\otimes$}}
\Gluon(145,20)(165, 37){2}{4}
\Gluon(145,20)(165, 5){2}{4}
\put(185, 20){+  148 terms;}
\end{picture}
\end{center}
\item {\bf real-virtual:} the interference of the one-loop and  the Born 
  amplitudes for $gg \rightarrow Hg$, $gq \rightarrow Hq$, and 
  $g\bar q \rightarrow H \bar q$,   
\begin{center}
\begin{picture}(300,40)(0,0)
\Gluon(40,5)(60,5){2}{4}
\Gluon(60,5)(60,35){2}{6}
\Gluon(40,35)(60,35){2}{4}
\Gluon(60,35)(100, 20){2}{8}
\Gluon(60,5)(100, 20){2}{8}
\Gluon(80,27.5)(115,35){2}{5}
\put(97,17){{ $\otimes$}}
\SetWidth{1}
\Line(106,20)(115,20)
\Line(130,20)(137,20)
\SetWidth{0.5}
\put(135,17){{ $\otimes$}}
\Gluon(145,20)(165, 35){2}{4}
\Gluon(145,20)(165, 5){2}{4}
\Gluon(157.5,12.5)(125,5){2}{5}
\put(185, 20){+  635 terms;}
\end{picture}
\end{center}
\item  {\bf double-real:} the self-interference of the Born amplitudes for 
  $gg \rightarrow Hgg$, 
  $gg \rightarrow Hq \bar q$, $gq \rightarrow Hgq$, $g\bar q \rightarrow Hg
  \bar q$, $qq \rightarrow Hqq$, and $q \bar q \rightarrow H q \bar q$, 
\begin{center}
\begin{picture}(300,37)(0,0)
\Gluon(40,5)(60,5){2}{4}
\Gluon(60,5)(60,35){2}{6}
\Gluon(40,35)(60,35){2}{4}
\Gluon(60,35)(100, 35){2}{8}
\Gluon(60,20)(100, 20){2}{8}
\put(58,3){{ $\otimes$}}
\SetWidth{1}
\Line(67,5)(100,5)
\Line(130,20)(137,20)
\SetWidth{0.5}
\put(135,18){{ $\otimes$}}
\Gluon(145,20)(165, 35){2}{4}
\Gluon(145,20)(165, 5){2}{4}
\Gluon(157.5,12.5)(125,5){2}{5}
\Gluon(157.5,27.5)(125,35){2}{5}
\put(185, 20){ +  594 terms.}
\end{picture}
\end{center}
\end{itemize}
The above interference terms are  produced in a form convenient for further 
evaluation using the QGRAF package~\cite{qgraf} for generating Feynman graphs.

In the following subsections we briefly describe the available techniques for 
evaluating virtual corrections and explain  our
method for integrating over the phase-space of the 
final state particles.   

\subsection{Virtual corrections}

There currently exists a general
method which permits  the systematic evaluation 
of multi-loop virtual corrections.
In order to calculate a multi-loop amplitude we must first reduce the 
number of  Feynman integrals.
The  hypergeometric structure of Feynman integrals guarantees  that  
simple algebraic relations between various scalar integrals exist, making
such a reduction possible.  
One method of producing these  relations is the integration by parts 
(IBP) technique~\cite{IBP}. 
In cases where the system of equations is not
complete, it can be  supplemented  with additional identities that exploit
the Lorentz invariance (LI) of scalar integrals~\cite{LI}.

In general, IBP and Lorentz Invariance (LI) identities relate integrals 
of differing complexity. For example, it is possible that a single IBP
equation relates an integral with an irreducible
scalar product to integrals with no irreducible scalar products  or 
to integrals with fewer  propagators. A typical situation 
however, involves multiple IBP and LI identities relating several equally  
complicated integrals to a set of simpler ones. 
In such cases, every integral must be written exclusively in terms of 
simpler  ones and, eventually,  expressed in terms of a few ``master'' 
integrals which cannot be reduced further.  Unfortunately, finding recursive 
solutions of the IBP and LI identities is tedious, and may be impossible 
in complicated cases. Also, a separate treatment of  
each different topology in a Feynman amplitude is required. 
Consequently, the whole  procedure becomes increasingly cumbersome with 
the introduction of more kinematic variables  and loops. 

We may  alternatively consider a sufficiently large system of
explicit IBP and LI equations which  contains all the  integrals that 
contribute to  the multi-loop amplitude of interest. 
It should then be possible to solve the system of equations in terms of 
the master integrals~\cite{LI,laporta} using standard 
linear algebra elimination algorithms. In this approach, the number of loops,
the topological details, and the number of kinematic variables, affect 
only the size of the system of equations and the number of terms in each of 
the equations; they have  no bearing on the construction of the 
elimination algorithm. This in principle allows us to express any 
multi-loop amplitude in terms of master integrals.

One possible elimination algorithm has been proposed by 
Laporta~\cite{laporta}.  This algorithm exploits the fact that Feynman 
integrals can be ordered by their complexity; for example they can be
arranged according to the number of irreducible  scalar products 
and the total number and powers of  propagators. 
This observation distinguishes the IBP and LI systems of 
equations from  algebraic systems with no intrinsic ordering,
and it becomes possible to solve them iteratively, starting with the simpler
equations and progressing  to
more complicated. We  use a
variant of this algorithm, implemented in FORM~\cite{FORM} 
and MAPLE~\cite{MAPLE}. 

After the reduction we must compute the analytic expansion in $\epsilon$ of
the master integrals. The coefficients of the expansion 
are  typically expressed in terms of polylogarithms whose  
rank and complexity depends  on  the number of loops and  kinematic 
variables of the integral in question.   The Mellin-Barnes 
representation~\cite{mellinbarnes} and  the differential equation 
method~\cite{LI} can be used to evaluate master integrals explicitly.

\subsection{Reduction of phase-space integrals}

In this subsection we  extend the application of the above techniques 
to calculate phase-space integrals for inclusive cross-sections. 
To the best of our knowledge  the method we present is new, however 
a somewhat related discussion has been given earlier in \cite{baikov}.

To illustrate our method,  we  consider 
the following double-real contribution at NNLO:
\begin{equation}
\label{eq:square}
\left|
\begin{picture}(60,35)(0,25)
\Gluon(12, 5)(40, 5){2}{6}
\Gluon(12, 35)(25, 35){2}{3}
\Gluon(25, 35)(25, 5){2}{5}
\Gluon(25, 20)(40, 20){2}{3}
\put(23,33){{\scriptsize { $\otimes$}}}
\put(0,35){{\scriptsize { $p_1$}}}
\put(0,5){{\scriptsize { $p_2$}}}
\put(42,35){{\scriptsize { $q_H$}}}
\put(42,20){{\scriptsize { $q_1$}}}
\put(42,5){{\scriptsize { $q_2$}}}
\SetWidth{1}
\Line(29,35)(40, 35)
\end{picture}
\right|^2
 \sim \int \frac{d^d q_1 d^d q_2
\delta(q_1^2) \delta(q_2^2) \delta(q_H^2 -m_H^2) \left[ \ldots \right]
}{\left[ (q_H-p_1)^2\right]^2 
\left[ (q_2-p_2)^2\right]^2 }.
\end{equation}
Using the Cutkosky rules~\cite{cutkosky}, we can replace the 
delta-functions in the above integral by differences of two propagators:
\begin{equation}
\label{eq:cutrule}
2 i \pi \delta \left(p^2-m^2\right) \rightarrow \frac{1}{p^2-m^2+i0} - \frac{1}{p^2-m^2-i0}. 
\end{equation}
The r.h.s. of Eq.~(\ref{eq:square}) is now equal to a forward
scattering diagram:
\begin{center}
\begin{equation}
\label{fig:cutrule}
\begin{picture}(220,35)(0,0)
\Gluon(10, 5)(25, 5){2}{3}
\Gluon(10, 25)(25, 25){2}{3}
\Gluon(25,5)(25,25){2}{3}
\Gluon(25,5)(40, 5){2}{3}
\Gluon(25,17.5)(40, 17.5){2}{3}
\Gluon(50,17.5)(65, 17.5){2}{3}
\Gluon(50,5)(65, 5){2}{3}
\Gluon(65, 5)(65, 25){2}{3}
\Gluon(65, 5)(80, 5){2}{3}
\Gluon(65, 25)(80, 25){2}{3}
\Gluon(140, 5)(200, 5){2}{14}
\Gluon(155, 18)(185, 18){2}{6}
\Gluon(155, 5)(155, 28){2}{4}
\Gluon(185, 5)(185, 28){2}{4}
\Gluon(140, 30)(155, 30){2}{3}
\Gluon(185, 30)(200, 30){2}{3}
\put(20,23){{\scriptsize { $\otimes$}}}
\put(59,23){{\scriptsize { $\otimes$}}}
\put(150,28){{\scriptsize { $\otimes$}}}
\put(180,28){{\scriptsize { $\otimes$}}}
\put(110, 15){$=$}
\put(205, 5){{\scriptsize $p_2$}}
\put(205, 30){{\scriptsize $p_1$}}
\put(125, 5){{\scriptsize $p_2$}}
\put(125, 30){{\scriptsize $p_1 \quad \quad ,$}}
\SetWidth{1}
\Line(28,25)(40, 25)
\Line(50,25)(63, 25)
\Line(158, 30)(182, 30)
\DashLine(170, 35)(170, 0){5}
\end{picture}
\end{equation}
\end{center}
where a cut  propagator should be replaced by the r.h.s. 
of Eq.(\ref{eq:cutrule}).

We have exchanged the square of a Born amplitude for a two-loop diagram, in
contrast to the usual application of the Cutkosky rules. We do this in order
to utilize IBP and LI relations between multi-loop integrals. The phase-space
integrals can then be evaluated in the same algorithmic fashion as the
multi-loop integrals. 

We begin our calculation by summing  over  the colors
and spins of the external particles in the cut two-loop integral on 
the right hand side of Eq.(\ref{fig:cutrule}). The original 
diagram is then expressed in terms of a large number of scalar two-loop integrals to which  the same cutting rules apply.  
Crucially, we can use the IBP method to reduce the cut scalar integrals. 
This is a consequence of the fact that the delta-function in
Eq.~(\ref{eq:cutrule}) is  represented in a very simple manner by the 
difference of two propagators with opposite prescriptions for their 
imaginary parts.  We derive the IBP equations by integrating over total 
derivatives which act on the propagators of the cut scalar integrals. The
prescription for the imaginary part of the two propagators in the r.h.s. of
Eq.~(\ref{eq:cutrule}) is irrelevant for the differentiation.
Therefore the IBP relations for the two descendants of these two terms
have the same form as the IBP relations for the original integral without 
the cut. It is then allowed to commute the application of IBP reduction 
algorithms with the application of the Cutkosky rules.
  
After the IBP reduction, the original phase-space integral 
is expressed in terms of a small number of master integrals cut through the 
same three propagators as the initial diagram\footnote{
Bold lines represent a massive Higgs propagator. Normal lines denote 
massless scalar propagators.}:  
\begin{center}
\begin{equation}
\begin{picture}(300, 40)(0, 0)
\Gluon(0,5)(60, 5){2}{12}
\Gluon(15, 5)(15, 35){2}{5}
\Gluon(45, 5)(45, 35){2}{5}
\Gluon(15, 20)(45, 20){2}{6}
\Gluon(0,35)(11,35){2}{2.0}
\Gluon(46,35)(60,35){2}{3}
\Line(95,5)(135, 5)
\Line(95,35)(100,35)
\Line(130,35)(135,35)
\Line(100,20)(130, 20)
\Line(100,5)(100,35)
\Line(130,5)(130,35)
\Line(190,5)(230, 5)
\Line(190,35)(195,35)
\Line(225,35)(230,35)
\Line(195,5)(225, 35)
\Line(195,5)(195,35)
\Line(225,5)(225,35)
\put(9,33){{\scriptsize { $\otimes$}}}
\put(39,33){{\scriptsize { $\otimes$}}}
\SetWidth{1}
\put(69, 20){{ $=$}}
\put(79, 20){{ $A_1$}}
\put(150, 20){{ $+ \;\;\;\;\; A_2$}}
\put(240, 20){{ $+\ldots \quad \quad .$}}
\SetWidth{2}
\Line(16,35)(41,35)
\Line(100,34.5)(130, 34.5)
\Line(195,34.5)(225, 34.5)
\SetWidth{1}
\DashLine(30, 40)(30, 0){2}
\DashLine(115, 40)(115, 0){2}
\DashLine(210, 40)(210, 0){2}
\end{picture}
\end{equation}
\end{center}
During the reduction, integrals with one or more of 
the cut propagators eliminated are produced. 
From Eq.~(\ref{eq:cutrule})  we observe
that such terms  do not contribute  to the  original phase-space
integrals. Therefore, we can  immediately discard them simplifying  
the reduction process.     

A similar procedure can be applied to the   virtual-real
contributions. In this case, since we perform the phase-space integration 
over two final state particles, the resulting master integrals should be
cut through two of the propagators:
\begin{equation}
\begin{picture}(80, 60)(0, 27)
\Gluon(0,5)(60, 5){2}{12}
\Gluon(0,50)(10, 50){2}{2}
\Gluon(32,50)(60, 50){2}{5}
\Gluon(10,48)(10,5){2}{8}
\Gluon(50,48)(50,5){2}{8}
\Gluon(33,49)(50,5){2}{8}
\put(8,48){{\scriptsize { $\otimes$}}}
\put(30,48){{\scriptsize { $\otimes$}}}
\DashLine(27, 60)(27, 0){2}
\SetWidth{2}
\Line(15, 50)(32, 50)
\end{picture}
= B_1 \YEb{27} + B_2 \YAb{27} + \ldots
\end{equation}
\\
\\
\\
In order to have a  unified algorithm for all three types of interferences, 
we treat the double-virtual corrections as integrals with a 
single cut through the propagator of the Higgs boson.  
\begin{equation}
\begin{picture}(90, 50)(0, 27)
\Gluon(0,5)(50, 30){2}{10}
\Gluon(0,55)(50,30){2}{10}
\Gluon(8,55)(8, 5){2}{6}
\Gluon(8,30)(50,30){2}{8}
\Gluon(72,32)(80,45){2}{2}
\Gluon(72,28)(80,15){2}{2}
\put(48,28){{\scriptsize { $\otimes$}}}
\put(67,28){{\scriptsize { $\otimes$}}}
\DashLine(63, 40)(63, 20){2}
\SetWidth{2}
\Line(55, 30)(70, 30)
\end{picture}
= C_1 \sunsb{27} + C_2 \glassb{27} + \ldots \quad \quad .
\end{equation}
\vspace{0.5cm}

Finally we must evaluate the master integrals
as a series expansion in $\epsilon=(4-d)/2$. Since each of the cut master
integrals represents a well-defined phase-space integral, we could  compute
them using brute force  techniques similar to the ones 
described in  Ref.~\cite{drellyan}.  However, we can instead utilize 
the IBP reduction algorithm in order to produce a set 
of coupled first order differential equations \cite{LI}
that the master integrals satisfy. It is simpler to solve the 
differential equations than to reinstate the delta-functions for the 
cut propagators and perform the integrations over the phase-space.

\subsection{Evaluation of phase-space master integrals}
To explain how the system of differential equations for the master integrals 
is obtained, let us consider a two-loop scalar integral with a single 
Higgs boson propagator
\begin{equation}
{\cal I}(s,\M) = \measure \frac{1}{\left[ k^2-\M \right]^{\nu} A_1^{\nu_1}
\ldots A_n^{\nu_n}} .
\end{equation}
By differentiating  with respect to $\M$ we obtain:
\begin{equation}
\frac{\partial {\cal I}(s,\M)}{\partial \M} 
= \nu \measure \frac{1}{\left[ k^2-\M
  \right]^{\nu+1} A_1^{\nu_1} \ldots A_n^{\nu_n}}.
\end{equation}
After applying  the IBP algorithm,  we can rewrite the r.h.s of the last
equation in  terms of the master integrals $\{ {\cal X}_i\}$, yielding
\begin{equation}
\label{eq:derivscalar}
\frac{\partial {\cal I}}{\partial \M} = \sum \limits_{j}^{}c_{j} {\cal X}_j.
\end{equation} 
By identifying ${\cal I} = {\cal X}_i$, 
 we derive a closed system of
differential equations for the master integrals, 
\begin{equation}
\label{eq:diffsystem}
\frac{\partial {\cal X}_i}{\partial \M} = \sum 
\limits_{j}^{}c_{ij} {\cal X}_j.
\end{equation}
These differential equations can be  solved up to a constant 
in terms of logarithms and generalized Nielsen polylogarithms, order by order
in $\epsilon$. This constant is obtained by 
evaluating  the master integral at a specific kinematic point. This is   
typically simpler and can often be avoided 
using general arguments, as shown in an example below. 

As an explicit example we discuss the calculation 
of the master integrals for the real-virtual corrections. 
The IBP reduction produces  six master integrals which 
depend on  two variables: the mass of the 
Higgs boson, $m_H$, and the square of the sum of the incoming momenta, 
$\hat s=(p_1+p_2)^2$. Note that the integrals depend on a single  Mandelstam 
variable, since they correspond to forward scattering 
diagrams with the same incoming and outgoing momenta. It is convenient to
express the master integrals in terms of  the dimensionless ratio
$z=\M/\hat s$. We can further simplify our results by setting $\hat s=1$. 
The full dependence on $\hat s$ can be restored by simple dimensional 
analysis.  

Three of these master integrals are combinations of the one-loop massless
bubble integral and the two-body phase-space integral and can be easily 
evaluated, yielding 
\vspace*{-2cm} \hfill \\
\begin{eqnarray}
\YE{27} 
&=& \PS \LOOP {\rm Re}\left( e^{-i\pi\epsilon}\right) (1-z)^{1-2\epsilon}, 
\label{m1}
\end{eqnarray}
\vspace*{-2cm} \hfill \\
\begin{eqnarray}
~~~~~~~~~\YA{27} 
&=& \PS \LOOP  \frac{1-2\epsilon}{1-3\epsilon}
\frac{\Gamma(1-2\epsilon)^2}{\Gamma(1-\epsilon)\Gamma(1-3\epsilon)} 
(1-z)^{1-3\epsilon}, 
\label{m2}
\end{eqnarray}
\vspace{-1cm} \hfill \\
\begin{eqnarray}
\YB{27}
&=& \PS \LOOP  {\rm Re} \left(e^{-i\pi\epsilon}\right) z^{-\epsilon} (1-z)^{1-2\epsilon},
\label{m3}
\end{eqnarray}
\vspace*{0.5cm}
where 
\begin{equation}
\PS = 
\frac{\pi^{\frac{3}{2}-\ep}}{\Gamma\left(\frac{3}{2}-\ep \right) 2^{3-2\ep}}, \quad
\quad 
 \LOOP = \frac{\Gamma(\ep) \Gamma(1-\ep)^2}{\Gamma(2-2\ep)}.
\end{equation}
The remaining three master integrals, 
\vspace*{-0.5cm}
\begin{equation}
\YC{27},~~~~~~~~~~~~~~~~ 
\YD{27},~~~~~~~~~~~~~~~
\YF{27}
\label{mcompl}
\end{equation}
\\
\\
have a more complicated dependence on $z$ and we compute them using the 
method of differential equations. As an example, we discuss 
the differential equation for 
the first integral in Eq.(\ref{mcompl}): 
\vspace*{-1cm}
\begin{eqnarray}
\left ( 
\frac{\partial }{ \partial z}
 + \frac{2 \epsilon}{z} \right ) \hspace*{-0.4cm}\YCb{27} \hspace*{-0.5cm}=
-\frac{(1-2\epsilon)^2}{\epsilon z (1-z)} \hspace*{-0.4cm} \YBb{27} 
\hspace*{-0.5cm}
+ \frac{(1-2\epsilon)(1-3\epsilon)}{\epsilon z (1-z)}  \hspace*{-0.4cm}\YAb{27}
\label{eq:exampleDE}
\end{eqnarray}
\\
This differential equation is of the form
\begin{equation}
\left ( \frac{\partial }{\partial z} - \alpha(z) \right ) f(z) 
= \beta(z), 
\end{equation}
and has the general solution
\begin{equation}
\label{eq:generalsol}
f(z)= e^{\int^z dx \alpha(x) } \left( C
+ \int_{0}^z dx \beta(x) e^{-\int^x dx^\prime \alpha (x^\prime)}   \right).
\end{equation}
Using  Eq.(\ref{eq:generalsol}) 
and the expressions for the two boundary master integrals in
Eqs.(\ref{m2},\ref{m3}), we derive
\vspace*{-1cm}
\begin{eqnarray}
\YCb{27}&=&z^{-2\ep} \frac{(1-2\ep)^2}{\ep}\Bigg\{
{\rm Re}\left( e^{i\pi\epsilon} \right) \int_0^z dx x^{-1-\ep} (1-x)^{-2\ep}
\nonumber \\
&& 
+\frac{\Gamma(1-2\ep)^2}{\Gamma(1-3\ep) \Gamma(1-\ep)} 
\int_0^z dx x^{-1-2\ep} (1-x)^{-3\ep} + {\cal C}
\Bigg\}. 
\label{eq:integralform}
\end{eqnarray}
We compute the value of  
the  integral at a specific kinematic  point in order to determine the constant ${\cal C}$. 
A convenient choice is the threshold for  Higgs production, $z=1$, where the 
integral vanishes:
\vspace*{-1cm}
\begin{equation}
\YCb{27} \hspace{-15pt} (z=1) \hspace{10pt} =  0.
\end{equation}
\\ 
The value of the integral at $z=1$ can be inferred without an explicit
calculation by observing that, in the $z \to 1$  limit  
the  two particle phase-space scales like $(1-z)$, and the one-loop triangle 
diagram on the r.h.s. of the cut scales like $(1-z)^0$.
Requiring that Eq.~(\ref{eq:integralform}) vanishes at $z=1$, one finds: 
\begin{equation}
C = - \left [
{\rm Re}\left( e^{i\pi\epsilon} \right) {\rm B}(-\epsilon,1-2\epsilon)
+\frac{\Gamma(1-2\epsilon)^2}{\Gamma(1-3\epsilon)\Gamma(1-\epsilon)}
{\rm B}(-2\epsilon,1-3\epsilon)
\right ].
\end{equation}
We now evaluate the integrals in Eq.(\ref{eq:integralform}). 
First, by changing the integration variables, we isolate the 
singularity at $x=0$:   
\begin{eqnarray}
\int_0^z dx x^{-1-a\ep} (1-x)^{-b\ep} &=&  z^{-a \ep} \left\{ 
-\frac{1}{a\ep} + \int_0^1 dy \; y^{-a \ep} \frac{(1-zy)^{-b \ep} -1}{y}
\right\}.  
\end{eqnarray}
The integrand on the r.h.s. can then be expanded in $\epsilon$, yielding 
generalized Nielsen polylogarithms, 
\begin{equation}
S_{np}(z) = \frac{(-1)^{n+p-1}}{(n-1)! p!} \int_0^1 dy \; \frac{\log^{n-1}(y)
\log^p(1-zy)}{y},
\label{eq32}
\end{equation}  
which reduce to usual polylogarithms 
for  $p=1$: 
\begin{equation}
{\rm S}_{n-1,1}(z) \equiv {\rm Li}_n(z).
\end{equation} 
We finally obtain
\begin{eqnarray}
\int_0^z dx x^{-1+a\ep} (1-x)^{b\ep} &=&  -\frac{z^{-a \ep}}{a\ep} \left\{ 
 1 - \sum_{n,p}^{\infty} a^n b^p \ep^{n+p} {\rm S}_{np}(z)
\right\}.  
\end{eqnarray}
Substituting this result in  Eq.(\ref{eq:integralform}) and truncating the 
series at
the order where polylogarithms of rank $n+p>3$ start to appear, we arrive at 
the result:
\vspace*{-1cm}
\begin{eqnarray}
\YCb{27} &=& \PS \Gamma(1+\ep) \frac{z^{-2\ep}}{1-2\ep} \Bigg\{ \frac{1}{\ep} \left[  
\lid{z} +\frac{\log^2(z)}{2} -\zd
\right] 
+ 5 \Sab{z} - 4 \lit{z} + 4 \log(z) \lid{z} 
\nonumber \\
&& \hspace{-1cm}
+\frac{\log^3(z)}{2} +2 \zd \log(z) -\zt
+\O{\ep} \Bigg\}.  
\end{eqnarray}
We repeat the same procedure for the 
differential equations for  the two  remaining integrals 
in Eq.(\ref{mcompl}), where
the master integral we have just calculated enters as a boundary term. It is
important to extract the singular behavior of the master integrals 
around $z=1$ before expanding in $\epsilon$.
This is essential since terms of the form 
$(1-z)^{-1+a\ep}$ in the cross-section 
are expanded in $\epsilon$ in terms of ``plus'' distributions, 
\begin{eqnarray}
(1-z)^{-1+a\ep} &=& \frac{1}{a\ep} \delta(1-z) 
+ a\ep \left[\frac{1}{1-x}\right]_{+}
+ \frac{(a\ep)^2}{2!} \left[\frac{\log(1-z)}{1-z}\right]_{+}
+ \frac{(a\ep)^3}{3!} \left[\frac{\log^2(1-z)}{1-z}\right]_{+}
+{\cal O}(\ep^4),
\end{eqnarray}
facilitating the cancellation between real and virtual 
soft and collinear  singularities prior to integration over $z$.

Finally, we apply the same technique to compute the double real master 
integrals. Explicit formulae for the required master integrals are given in
the Appendix. 


\section{Partonic cross-sections}
\label{sec:results}
In this section we present analytic expressions for the partonic
cross-sections $i + j \to H + X$ of Eq.~(\ref{eq14}).  We write
\be
\hat \sigma_{ij} = \sigma_0 \left [ \eta_{ij}^{(0)} +
\left (\frac{\alpha_s}{\pi} \right ) \eta_{ij}^{(1)} + 
\left (\frac{\alpha_s}{\pi} \right )^2 \eta_{ij}^{(2)}
+{\cal O}(\alpha_s^3)
   \right],
\label{eq:partonicexpansion}
\ee
where 
$$
\sigma_0 = \frac{\pi}{576 v^2} \left ( \frac{\alpha_s}{\pi} \right)^2,
$$
and  $\alpha_s$ is the $\overline {\rm MS}$ strong coupling constant evaluated 
at the scale $\mu_r = m_H$. For simplicity, the factorization scale 
is also set equal to the mass of the Higgs boson $\mu_f=m_H$.

At leading order we find
\ba
\eta^{(0)}_{ij} = \delta(1-x)\delta_{ig}\delta_{jg}.
\ea

At next-to-leading order there are contributions from the gluon-gluon,
quark-gluon and quark-antiquark channels:
\ba
&&\eta^{(1)}_{gg} = 
\left ( \frac{11}{2} + 6 \zeta_2   \right ) \delta(1-x) 
+ 12 \left [ \frac{\ln(1-x)}{1-x} \right ]_+
- 12x(-x+x^2+2)\ln(1-x)
\nonumber \\
&& 
-\frac{6 (x^2+1-x)^2}{1-x}\ln(x)
-\frac{11}{2} (1-x)^3,
\\
&&\eta^{(1)}_{qg}  = 
-\frac{2}{3} \left ( 1+(1-x)^2 \right )\ln \frac{x}{(1-x)^2}-1+2x
-\frac{1}{3}x^2,
~~~~~~~~~~\eta^{(1)}_{q \bar q}  = \frac{32}{27}(1-x)^3.
\ea
The main result of this paper is the next-to-next-to-leading order 
corrections, which we separate according to their dependence on the number 
of quark flavors: 
\be
\eta_{ij}^{(2)} = \Delta_{ij}^{(2){\rm A}} + n_f \Delta_{ij}^{(2){\rm F}}.
\ee

For the gluon-gluon channel we find 
\ba
&& \Delta_{gg}^{(2){\rm A}} = 
\left ( \frac{11399}{144}+\frac{133}{2}\zeta_2 -\frac{165}{4}\zeta_3
-\frac{9}{20} \zeta_2^2+\frac{19}{8}L_t \right )\delta(1-x)
+\left( 133- 90 \zeta_2 \right) \left[\frac{\ln(1-x)}{(1-x)}\right]_+
\nonumber \\
&& +\left ( -\frac{101}{3}+33 \zeta_2+ \frac{351}{2}\zeta_3 \right)
 \left[\frac{1}{(1-x)} \right]_+
-33 \left[ \frac{\ln(1-x)^2}{(1-x)} \right ]_+
+72 \left [ \frac{\ln(1-x)^3}{(1-x)} \right ]_+
\nonumber \\
&& + \frac{9 (38 x^2-20 x^3+18 x-39 x^4+14+7 x^5)}{1-x^2}\Li_3(x)
- \frac{18 (x^2+x+1)^2}{1+x}{\rm S}_{12}(x^2)
\nonumber \\
&& + \frac{9 (4 x^4+8 x^3+21 x^2+14 x+7)}{1+x} {\rm S}_{12}(-x)
- \frac{9}{2} \frac{(5 x^5-51 x^4-57 x^3+53 x^2+59 x-11)}{1-x^2}{\rm S}_{12}(x)
\nonumber \\
&&
 - \frac{9}{2} \frac{(8 x^4+8 x^3-3 x^2-2 x-1)}{1+x}\Li_3(-x)
-\frac{9}{2} \frac{(16+13 x^5-40 x^3-67 x^4+64 x^2+36 x)}{1-x^2}\Li_2(x)\ln(x)
\nonumber \\
&&
 +\frac{9}{2} \frac{(2 x^4-15 x^2-10x-5)}{1+x}\Li_2(-x)\ln(x)
 -\frac{9}{4} \frac{(59+177 x^2-116 x^3+59 x^4-118 x)}{1-x}\ln(x)\ln^2(1-x)
\nonumber \\
&& 
+ \frac{27 (3 x^2+2 x+1)}{1+x} \Li_2(-x)\ln(1+x)
  +\frac{9 (6-11 x^3+18 x^2-12 x+6 x^4)}{1-x}\ln^2(x)\ln(1-x)
\nonumber \\
&&  +\frac{9}{2}\frac{(3-8 x^3+3 x^4-6 x+9 x^2)}{1-x}\Li_2(x)\ln(1-x)
 -\frac{3}{2} \frac{(7 x-7 x^3+4+18 x^2-17 x^4+9 x^5)}{1-x^2}\ln^3 (x)
\nonumber \\
&& 
+ \frac{9}{2} \frac{(8x^4+16x^3+33x^2+22x+11)}{1+x}\zeta_2 \ln(1+x)
-\frac{36 (x^2+x+1)^2}{1+x} \Li_2(x) \ln(1+x)
\nonumber \\
&& -\frac{9}{4}\frac{(4 x^4+8 x^3+27 x^2+18 x+9)}{1+x}\ln(1+x)\ln^2 (x)
+(-21+\frac{63}{2} x^2-18 x+\frac{33}{2} x^3) \ln(1+x)\ln(x)
\nonumber \\
&&
+\frac{27}{2} \frac{(3 x^2+2 x+1)}{1+x}\ln^2(1+x)\ln(x)
-\frac{3}{4} \frac{(-280 x^3+143 x^4+394 x-289+21 x^2)}{(1-x)}\Li_2(x)
\nonumber \\
&& 
  +(-21+\frac{63}{2} x^2-18 x+\frac{33}{2} x^3)\Li_2(-x)
+(-\frac{2559}{4} x^3+\frac{1079}{2} x^2-\frac{2687}{4} x+\frac{2027}{4})
\ln(1-x)
\nonumber \\
&& 
-\frac{3}{8} \frac{(374 x^4-389 x+154+699 x^2-827 x^3)}{1-x}\ln^2(x)
 +(330 x^3-348 x^2+381 x-297)\ln^2(1-x)
\nonumber \\
&& 
+\frac{3}{4} \frac{(-1180 x^3+641-1238x+1227x^2+605x^4)}{1-x}\ln(x)\ln(1-x)
-72 (2-x+x^2) x \ln^3(1-x)
\nonumber \\
&& -\frac{1}{8} \frac{(4318x^4-6955x^3+6447x^2-5611x+2333)}{1-x}\ln(x)
+\frac{3}{4}\frac{(495x^4-886x^3+564x^2-200x+16)}{1-x}\zeta_2
\nonumber \\
&& +\frac{9 (6 x+18 x^2+2+10 x^5-6 x^3-19 x^4)}{1-x^2}\zeta_2 \ln(x)
-\frac{9}{2} \frac{(-48 x^3+23 x^4-46 x+3+69 x^2)}{1-x}\zeta_2 \ln(1-x)
\nonumber \\
&& +\frac{9}{2} \frac{(-36-15 x^4-52 x+19 x^2+13 x^3+33 x^5)}{1-x^2}\zeta_3
+ \frac{7539}{16}x^3-\frac{24107}{48} x^2+\frac{22879}{48} x-\frac{18157}{48},
\ea
and 
\ba
&& \Delta_{gg}^{(2){\rm F}} = 
( -\frac{1189}{144}+ \frac{5}{6}\zeta_3-\frac{5}{3}\zeta_2
 +\frac{2}{3}L_t )\delta(1-x)
-\frac{10}{3} \left[ \frac{\ln(1-x)}{1-x} \right]_+
+\left (\frac{14}{9}-2 \zeta_2 \right )\left [ \frac{1}{1-x} \right]_+
\nonumber \\
&& 
+ 2 \left [ \frac{\ln(1-x)^2}{1-x} \right ]_+
 + \left ( \frac{31}{6} x+ \frac{1}{6}+ \frac{65}{12}x^2 \right ) 
{\rm S}_{12}(x)
+ \left ( -\frac{31}{12} x^2+ \frac{1}{6}- \frac{17}{6} x \right )
\Li_3(x)
\nonumber \\
&&
+\left ( \frac{47}{12} x^2+ \frac{25}{6} x- \frac{1}{6} \right )
\Li_2(x) \ln(x)
+\left (-\frac{1}{12} x^2+\frac{1}{6} x-\frac{1}{6} \right) \zeta_2 \ln(1-x)
-4 x (1+x) \zeta_2 \ln(x)
\nonumber \\
&& + \left(-\frac{1}{6} x + \frac{1}{6}+\frac{1}{12} x^2 \right ) 
\Li_2(x)\ln(1-x)
+ \left( \frac{1}{12}- \frac{1}{12} x+ \frac{1}{24} x^2 \right )
\ln(1-x)\ln(x)\ln \left ( \frac{(1-x)}{x}\right )
\nonumber \\
&& +\frac {5}{9} x (1+x) \ln^3x
+\left (-\frac{17}{6} x^2-\frac{7}{3} x-\frac{1}{3} \right ) \zeta_3
+\left (-\frac{34}{9} x^3+ \frac{2}{3} x^2- \frac{8}{3} x+ 
\frac{16}{9} \right) \left ( \ln^2(1-x) - \zeta_2 \right )
\nonumber \\
&& 
-\frac{2}{9}\frac{(21 x^2+7 x+25 x^4+17-61 x^3)}{1-x}\ln(x)\ln(1-x)
+ \left ( \frac{785}{54}x^3-\frac{83}{36}x^2+\frac{49}{18}x-\frac{461}{54}
\right ) \ln(1-x)
\nonumber \\
&&
+\frac{1}{72}\frac{(-351x^3+117x^2+68+132x^4+52x)}{1-x}\ln^2(x)
+\frac{1}{36}\frac{(227x^3+68+4x^4-302x+21x^2)}{1-x}\Li_2(1-x)
\nonumber \\
&& 
+\frac{1}{216}\frac{(333x^2+2384x^4-598x-3041x^3+1282)}{1-x}\ln(x)
-\frac{8887}{648}x^3
+\frac{1267}{432}x^2
-\frac{497}{216}x+\frac{12923}{1296}.
\ea

For the quark-gluon channel we obtain
\ba
&& \Delta_{qg}^{(2)A} = 
\left ( \frac{170}{3} x+\frac{338}{9}+\frac{119}{3} x^2 \right )\Li_3(x)
+(4 x+4+2 x^2)\Li_3(-x)
+(16+8 x^2+16 x) S_{12}(-x)
\nonumber \\
&& + \left (-\frac{614}{9}x- \frac{269}{9}x^2
-\frac{74}{9} \right ) S_{12}(x)
+(-2 x^2-4-4 x) S_{12}(x^2)+\left ( \frac{367}{27}+\frac{367}{54} x^2
-\frac{367}{27} x \right )\ln^3(1-x)
\nonumber \\
&& +\left ( (2+x^2-2 x)\ln(1-x)-\left ( \frac{446}{9} x+ \frac{214}{9}
+\frac{281}{9} x^2 \right )\ln(x)
-(8+4 x^2+8 x) \ln(1+x) \right )\Li_2(x)
\nonumber \\
&& 
+(8 + 8 x + 4 x^2)\ln \left ( \frac{(1+x)}{x} \right ) \Li_2(-x)
+\left ( - \frac{115}{9} x^2+ \frac{230}{9} x- \frac{230}{9} 
\right )\ln(x)\ln^2(1-x)
\nonumber \\
&& + \left ( \frac{107}{9}+ \frac{107}{18}x^2
-\frac{107}{9}x \right )\ln^2(x)\ln(1-x)
+ \left ( - \frac{145}{54}x^2-\frac{71}{27} x-2 \right )\ln(x)^3
\nonumber \\
&& +(-3 x^2-6-6 x)\ln(1+x)\ln(x)^2+(4x+4+2x^2)\ln(1+x)^2\ln(x)
\nonumber \\
&& +\left (-\frac{4}{27}x^3-\frac{74}{9}x-\frac{11}{9}x^2
-\frac{166}{27} \right )\Li_2(-x)
 + \left ( \frac{2605}{54}- \frac{146}{9}x+\frac{74}{27} x^3
-\frac{79}{6}x^2 \right )\Li_2(x)
\nonumber \\
&&+\left ( \frac{1139}{18}x+\frac{37}{12}x^2+8x^3-72 \right )\ln^2(1-x)
+\left (- \frac{121}{18}x^2-\frac{326}{27}x^3-\frac{826}{9}x
+\frac{5935}{54} \right )\ln(x)\ln(1-x)
\nonumber \\
&&+\left ( \frac{113}{27}x^3+\frac{244}{9}x-\frac{13}{3}x^2-\frac{31}{2}
\right )\ln^2(x)
+ \left ( - \frac{4}{27}x^3-\frac{74}{9}x-\frac{11}{9}x^2-\frac{166}{27}
\right )\ln(1+x)\ln(x)
\nonumber \\
&& +\zeta_2 \left ( - \frac{59}{9}x^2+\frac{118}{9}x-\frac{118}{9}
\right )\ln(1-x)
+\zeta_2\left ( \frac{140}{9} x+ \frac{128}{9} x^2+ \frac{52}{9} \right )\ln(x)
\nonumber \\
&& +\zeta_2(12+12 x+6 x^2)\ln(1+x)
+ \left ( - \frac{392}{81} x^3-\frac{49}{3} x^2+ \frac{23671}{162}
-106 x \right )\ln(1-x)
\nonumber \\
&& + \left ( \frac{1985}{108} x^2+ \frac{800}{9} x- \frac{12209}{162}
+ \frac{616}{81} x^3 \right ) \ln(x)
+ \left ( - \frac{292}{27} x^3- \frac{82}{3} x+ \frac{16}{3} x^2
+\frac{221}{27} \right )\zeta_2
\nonumber \\
&& + \left (-18 x+10 x^2+ \frac{92}{9} \right )\zeta_3
-\frac{210115}{1944}+ \frac{1537}{486} x^3+
\frac{16465}{162} x+ \frac{2393}{648} x^2,
\ea
and
\ba
&& \Delta_{qg}^{(2)F} = 
\left ( \frac{1}{18} x^2-\frac{1}{9} x+\frac{1}{9} \right )\ln^2(1-x)
+ \left (- \frac{38}{27} x+\frac{19}{27} x^2+ \frac{29}{27} \right )\ln(x)
 - \frac{209}{81} x+ \frac{265}{162}
\nonumber \\
&& + \left (  \left ( - \frac{4}{9}+ \frac{4}{9} x
- \frac{2}{9} x^2 \right)\ln(x)- x^2+ \frac{16}{9} x
- \frac{13}{9} \right )\ln(1-x)
+ \frac{179}{162} x^2+ \left ( \frac{1}{9} x^2- \frac{2}{9} x
+ \frac{2}{9} \right )\ln^2(x).
\ea

For the scattering of two identical quarks we obtain:
\ba
&& \Delta_{qq}^{(2)A} = 
 \left ( \frac{368}{27} x+ \frac{104}{27}x^2
+\frac{400}{27} \right )\Li_3(x)
 -\frac{32}{9}(x+2)^2 S_{12}(x)
 -\frac{4}{27} (2+x^2-2x)\ln^2(x) \ln(1-x)
\nonumber \\
&& 
-\frac{4}{27} (x+2)^2\ln^3(x)
-\frac{16}{27} (19+5 x^2+17 x)\Li_2(x)\ln(x)
 -\frac{32}{9} (x+3) (1-x)\ln^2(1-x)
\nonumber \\
&& 
+ \frac{16}{3}(x+3)(1-x)\ln(x)\ln(1-x)
 + \frac{4}{27} (26 x-18+9 x^2)\ln^2(x)
- \frac{8}{9} (-6+x^2+4 x)\Li_2(x)
\nonumber \\
&& 
 +\frac{4}{3}(5x+17)(1-x)\ln(1-x)
+\left ( \frac{8}{9}(x+2)^2\zeta_2- \frac{118}{9}+ \frac{248}{27} x
+ \frac{46}{9}x^2 \right )\ln(x)
\nonumber \\
&& 
 + \left( -\frac{8}{27} x^2- \frac{16}{27}+ \frac{16}{27} x \right )\zeta_3
 + \left ( \frac{16}{3}- \frac{32}{9} x - \frac{8}{3} x^2 \right )\zeta_2
 - \frac{4}{27}(27x+160)(1-x),
\ea
and
\be
\Delta_{qq}^{(2)F}=0.
\ee
For the scattering of distinct quarks we find
\ba
&& \Delta_{qq'}^{(2)A} = 
\frac{32}{9} (x+2)^2 \left (\Li_3(x)- S_{12}(x) \right )
 - \frac{8}{3}(x+2)^2\ln(x)\Li_2(x) - \frac{4}{27}(x+2)^2\ln^3(x)
\nonumber \\
&& 
 - \frac{8}{9} (4 x-6+x^2)\Li_2(x) 
- \frac{32}{9}(x+3)(1-x)\ln^2(1-x)
+ \frac{16}{3}(x+3)(1-x)\ln(x)\ln(1-x)
\nonumber \\
&& 
+ \frac{8}{9}(x^2+4x-3)\ln^2(x)
 + \frac{8}{9}\zeta_2 (x+2)^2\ln(x)
+\frac{4}{3}(5x+17)(1-x)\ln(1-x)
\nonumber \\
&& 
+ \frac{2}{9}(29x^2+44x-59)\ln(x)
 + \left ( \frac{16}{3}-\frac{32}{9}x-\frac{8}{3}x^2 \right )\zeta_2
- \frac{2}{9}  ( 11x+105)(1-x),
\ea
and
\be
\Delta_{qq'}^{(2)F}=0.
\ee
Finally, for the quark-antiquark channel the NNLO contribution is
\ba
&& \Delta_{q \bar q}^{(2)A} = 
\left (- \frac{16}{9}-\frac{16}{9}x-\frac{8}{9}x^2 \right )\Li_3(-x)
 + \left ( - \frac{16}{27}x^2-\frac{32}{27}-\frac{32}{27}x\right )S_{12}(-x)
\nonumber \\
&&  + \frac{32}{9}(x+2)^2\Li_3(x)
 -\frac{32}{9}(x+2)^2S_{12}(x)
 -\frac{4}{27}(x+2)^2\ln^3(x)
+\frac{4}{9}(2+2x+x^2)\ln(1+x)\ln^2(x)
\nonumber \\
&& + \left ( - \frac{8}{27}(2+2x+x^2)\ln^2(1+x)
-\frac{8}{3}(x+2)^2\Li_2(x)
+\frac{8}{9}(2+2x+x^2)\Li_2(-x) \right )\ln(x)
\nonumber \\
&& -\frac{16}{27}(2+2x+x^2)\Li_2(-x)\ln(1+x)
 +\frac{32}{81}(1-x)(13x^2-35x-14)\ln^2(1-x)
\nonumber \\
&& 
-\frac{16}{81}(1-x)(37x^2-101x-44)\ln(x)\ln(1-x)
-\frac{8}{81}(44x^3+39x-81x^2+27)\ln^2(x)
\nonumber \\
&& 
+\frac{16}{27}x(x+6x^2+2)\ln(1+x)\ln(x)
+ \frac{8}{81}(42x-87x^2+12+10x^3)\Li_2(x)
\nonumber \\
&& 
+ \frac{16}{27}x(x+6x^2+2)\Li_2(-x)
- \frac{4}{81}(1-x)(384x^2-967x-75)\ln(1-x)
+ \left ( - \frac{16}{27}x^2-\frac{32}{27}-\frac{32}{27}x \right )\zeta_3
\nonumber \\
&& +\left ( \frac{8}{9}(x+2)^2\zeta_2+ \frac{4222}{81}x^2-\frac{2896}{81}x
-\frac{512}{27}x^3-\frac{10}{3} \right )\ln(x)
- \frac{8}{27}(2+2x+x^2)\zeta_2\ln(1+x)
\nonumber \\
&& 
+ \left ( \frac{752}{81}x^3- \frac{544}{27}x^2
+\frac{80}{81}+ \frac{400}{27}x \right )\zeta_2
+\frac{4}{81}(1-x)(783x^2-1925x+373),
\ea
and
\ba
\Delta_{q \bar q}^{(2)F} &=& 
\frac{32}{81}(1-x)^3\ln(1-x)
+ \left ( - \frac{64}{27}x^2+\frac{64}{81}x^3-\frac{16}{27}
+\frac{80}{27}x \right )\ln(x)
- \frac{8}{243}(1-x)(41x^2-88x+23).
\ea

The above results are valid if  the renormalization and
factorization scales are equal to the mass of the Higgs boson, 
$\mu_r = \mu_f= m_H$. It is easy to restore the complete functional dependence 
of the partonic cross-sections on these scales using the fact that 
the total hadronic cross-section is independent of them. 

We first find the dependence of the partonic cross-sections on 
a  scale $\mu$ which is equal  to both the  renormalization  and 
factorization scales $\mu_r = \mu_f = \mu$.
To do so, we  restore the full $\mu$ dependence in Eq.(\ref{eq14}):
\be
\sigma_{h_1 + h_2 \to H + X} = x \sum \limits_{ij} \left [
\bar f_i^{(h_1)}(\mu) \otimes \bar f_j^{(h_2)} (\mu) \otimes 
\left ( \hat  \sigma_{ij}(z,\mu) /z \right )
\right ](x).
\ee
Since the physical cross-section $\sigma_{h_1 + h_2 \to H + X}$
does not depend on $\mu$, 
\be
     \mu^2 \frac{d}{d \mu^2} \sigma_{h_1 + h_2 \to H + X} =0,
\ee
and the derivatives of the structure functions $\bar f_i^{(h)}$ with respect to
$\mu$ can be determined from the DGLAP evolution equation:
\be
\mu^2 \frac{{\rm d}}{{\rm d} \mu^2} \bar f_i (x, \mu) = \frac{\alpha_s(\mu)}{\pi}
\left [ P_{ij} \otimes \bar f_j(z,\mu) \right ](x), 
\ee
we derive the relation
\ba
0 && = \sum \limits_{ijk} \bar f_i^{(h_1)} \otimes 
\left [ 
\frac{\alpha_s(\mu)}{\pi}
P_{ik} \otimes \left ( \hat  \sigma_{kj}(z,\mu) /z \right )
+ \mu^2 \frac{{\rm d}}{{\rm d} \mu^2} \left ( \hat  \sigma_{ij}(z,\mu) /z \right )
+ \left ( \hat  \sigma_{ik}(z,\mu) /z \right )
\frac{\alpha_s(\mu)}{\pi}\otimes
P_{kj}  
\right ] \otimes \bar f_j^{(h_2)}.
\ea
This  equation  should hold for arbitrary $\mu$ and $x$; 
therefore, the expression in the 
square brackets should be identically zero for any choice of $i$ and $j$, 
yielding the following ``evolution equation'' for the cross-section:
\be
\mu \frac{{\rm d}}{{\rm d} \mu} \left ( \hat  \sigma_{ij}(z,\mu) /z \right )
= 
- \frac{\alpha_s(\mu)}{\pi} \Bigg[
P_{ik} \otimes \left ( \hat  \sigma_{kj}(z,\mu) /z \right )
 + \left ( \hat  \sigma_{ik}(z,\mu) /z \right ) \otimes
P_{kj}\Bigg].  
\label{eqmuf}
\ee
We can solve 
Eq.(\ref{eqmuf}) order by order in $\alpha_s(\mu)$ 
using the partonic cross-sections at $\mu = m_H$ as the boundary condition.  
Having obtained the explicit dependence of the partonic cross-sections on
$\mu$,
\be
\hat \sigma_{ij}(\mu) = \sigma_0 \left [ \eta_{ij}^{(0)}(\mu) +
\left (\frac{\alpha_s(\mu)}{\pi} \right ) \eta_{ij}^{(1)}(\mu) + 
\left (\frac{\alpha_s}{\pi}(\mu) \right )^2 \eta_{ij}^{(2)}(\mu)
+{\cal O}(\alpha_s^3)
   \right],
\label{eq:mupartonicexpansion}
\ee
we can find their dependence on two independent renormalization and 
factorization scales by expressing the 
strong coupling constant $\alpha_s(\mu = \mu_f)$ 
through $\alpha_s(\mu_r)$.

We have checked that our expressions for the  partonic cross-sections, 
derived by explicitly evaluating the Feynman amplitudes with their 
full scale dependence, are in agreement with Eq.~(\ref{eqmuf}). 
Our results are also in complete agreement with Ref.~\cite{hkhard}
 where the first sixteen terms of an expansion of the partonic 
cross-sections in $1-x$ were computed. In the limit, $x \to 0$, only the 
leading logarithmic corrections are known~\cite{hautmann}. We can easily 
reproduce this result by expanding our formulae for partonic cross-sections 
around $x=0$.

\section{Numerical results}
\label{sec:numerics}
We can now discuss the numerical impact of 
the NNLO corrections on the Higgs boson production cross-section 
at the LHC and the Tevatron. To calculate the cross-section we must convolute 
the hard scattering partonic cross-sections of Section~\ref{sec:results} 
with the appropriate parton distribution functions. For a self-consistent  
calculation at NNLO, we need the parton distribution functions 
at a given factorization scale at the same order.
At present, the NNLO evolution of the distribution functions 
can not be performed  since  
the required three-loop splitting functions are
not  known. Nevertheless, a significant number of moments of 
the splitting functions is available \cite{larin}, 
and this information can be combined   
with the known behavior at small $x$~\cite{smallx}, 
to obtain a useful approximation for the NNLO  
splitting functions~\cite{neervenvogt}. 
In Ref.~\cite{msrt} this approach was used to 
determine the NNLO MRST parton 
distribution functions; we  use these for the numerical evaluation 
of the Higgs boson production cross-section. 

\begin{figure}[h]
\hspace*{1 mm}
\begin{minipage}{16.cm}
\begin{picture}(100,0)
\put (150,150) {$\sigma(p p \to H + X)~[{\rm pb}],~~\sqrt{s} = 14~{\rm TeV}$}
\put (220,-10) {$m_H,~{\rm GeV}$}
\end{picture}
\psfig{figure=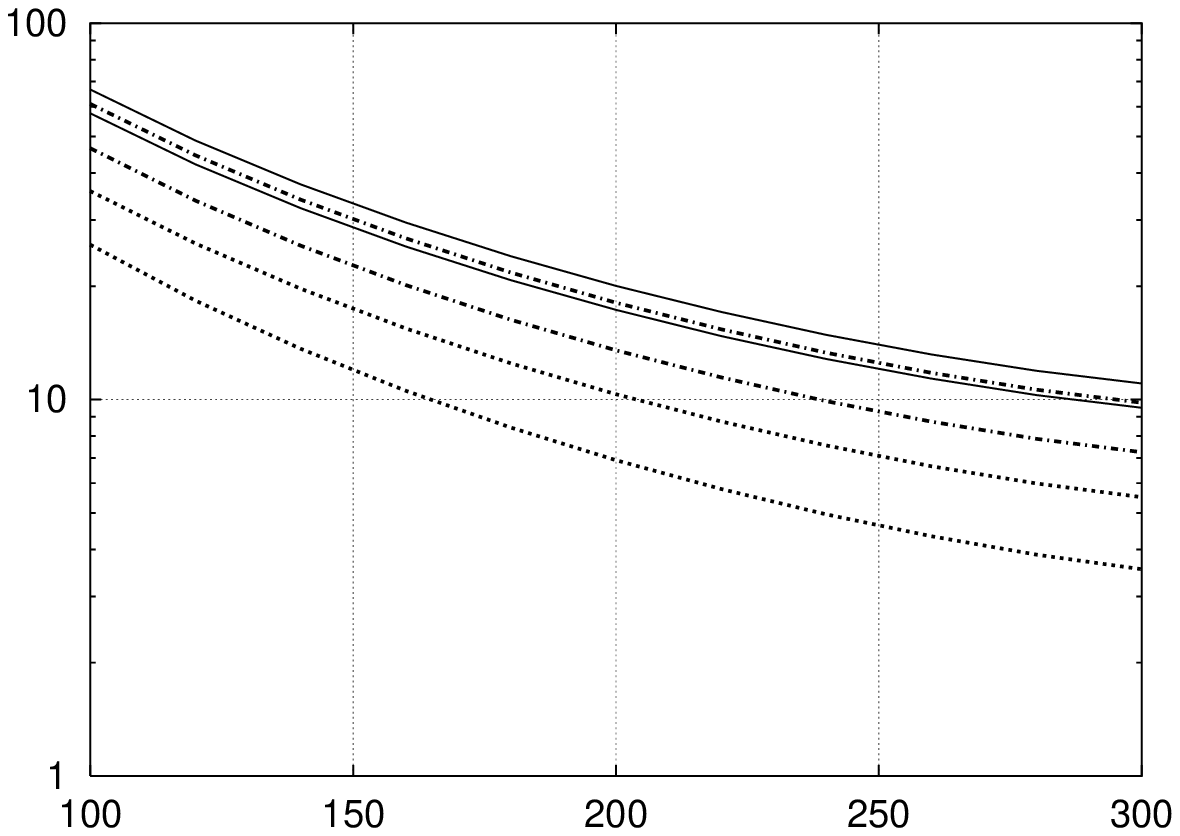,width=78mm}
\end{minipage}
\vspace*{0.5cm}
\caption{The Higgs boson production cross-section at the LHC at 
leading (dotted), next-to-leading (dashed-dotted) and
 next-to-next-to-leading (solid)
order. The two curves for each case correspond to $\mu_r = \mu_f = m_H/2$ (upper) and $\mu_r = \mu_f = 2 m_H$ (lower).}
\label{fig:plot1}
\end{figure}

To demonstrate the convergence properties of the perturbative 
series for the  hadronic cross-section, we present the LO, NLO and 
NNLO results for both the LHC and the Tevatron.  
In order to improve on the heavy top-quark approximation, we normalize the cross-section 
to the exact Born cross-section with the full $m_t$ dependence. 
We use  
the {\sf mode = 1} parton distribution functions (see Ref.~\cite{msrt} for
the notation).
For the NNLO set, 
this mode provides the ``average'' of two extreme cases, the so-called 
fast and slow evolutions. For the evaluation of the strong coupling constant
 we use  LO, NLO and NNLO running accordingly, with the $Z$-pole values 
used in the parton distribution functions as initial conditions 
(see \cite{msrt} for details). 
The total cross-section for the LHC is shown in Fig.\ref{fig:plot1}.
We note that the  NNLO cross-section does not vary significantly 
if we choose a different {\sf mode} for the MRST parton distribution
functions; the observed changes are less than $1\%$.

\begin{figure}[htb]
\hspace*{1 mm}
\begin{minipage}{16.cm}
\begin{picture}(100,0)
\put (150,150) {$\sigma(p \bar p \to H + X)~[{\rm pb}],~~\sqrt{s} 
 = 2~{\rm TeV}$}
\put (220,-10) {$m_H,~{\rm GeV}$}
\end{picture}
\psfig{figure=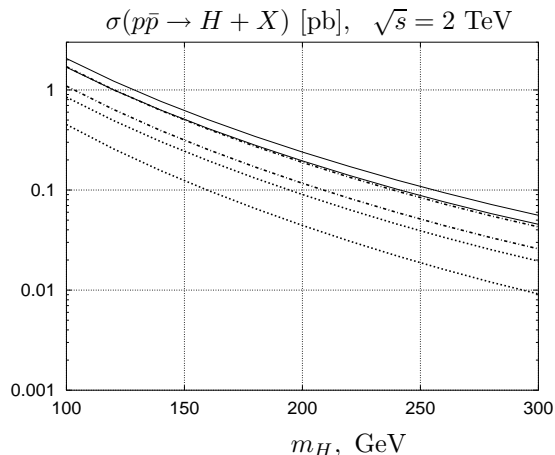,width=78mm}
\end{minipage}
\vspace*{0.5cm}
\caption{The Higgs boson production cross-section at the Tevatron at 
leading (dotted), next-to-leading (dashed-dotted) and 
next-to-next-to-leading (solid) order. The two curves for each case correspond to $\mu_r = \mu_f = m_H/2$ (upper) and $\mu_r = \mu_f = 2 m_H$ (lower).}
\label{fig:plot3}
\end{figure}

From Fig.\ref{fig:plot1} we observe that the scale dependence 
of the Higgs production cross-section at NNLO in the range 
$\mu_r=\mu_f = m_H/2 - 2 m_H$
is approximately 
$15\%$; this is a factor of two smaller than the NLO scale dependence 
and a factor of four smaller than the LO variation. Despite the scale
stabilization, the corrections are rather large;
the NLO corrections increase the LO cross-section by about $70\%$, 
and the NNLO corrections increase it further by approximately 
$30\%$. The $K$ factor, defined as the ratio of the NNLO cross 
section and the LO cross-section at $\mu_r=\mu_f=m_H$, is  approximately two.
In Fig.\ref{fig:plot3} we plot the values of the Higgs production 
cross-section at the Tevatron. The NNLO $K$ factor is approximately 
three, and  the residual scale dependence is approximately $23\%$. 
These numerical results for the NNLO Higgs boson production 
cross-section are in excellent agreement with the corresponding results
in \cite{hkhard}.

A few remarks concerning the magnitude of the corrections are appropriate. 
Despite the fact that the mass of the Higgs 
boson is much smaller than  the total center of mass energy, 
the production cross-section is dominated 
by partonic processes with $\hat s \sim m_H^2$,
This is because the gluon  gluon luminosity is a 
rapidly decreasing function of the partonic center of mass energy. 
The agreement between our numerical results based on the complete 
expressions of Section~\ref{sec:results} with the approximate results 
of~\cite{hkhard}, where an expansion in $1-\M/{\hat s}$ was employed,
demonstrates this indirectly.

The dominance of the threshold region renders resummation methods 
applicable~\cite{laenen,catanithr}.
However, threshold dominance should also affect the 
cross-section estimates based on fixed order 
calculations where there is freedom in the choice of the 
factorization scale. Since the production 
process is dominated by the region $x \to 1$, the appropriate 
factorization scale should be {\it parametrically  smaller} than 
the mass of the Higgs boson;  choosing a factorization 
scale near the Higgs boson mass may not capture the
essential physics of the process. 

We illustrate this point by considering the NLO correction to the 
Higgs production cross-section. Concentrating on the gluon-gluon subprocess, 
and keeping the most singular terms in the $x \to 1$ limit,
we can write
\be
\eta^{(1)}_{gg}(x) = \left ( \frac{\alpha_s}{\pi} \right ) \left \{
\left ( \frac{11}{2} + 6\zeta_2 \right ) \delta(1-x)  
-6 \left [ \frac{1}{1-x} 
\ln \left ( \frac{\mu^2}{m_H^2(1-x)^2} \right ) \right ]_+ + ... \right \}.
\ee
It is obvious from the above expression that if the dominant contribution 
to the integrated cross-section comes from the region 
$ x \sim 1$, then  choosing $\mu=m_H$ 
leaves large  logarithmic corrections of the form $\log(1-x)$ 
in the hard scattering cross-section. 
To avoid this problem, we should choose $\mu \sim m_H (1-x) $, 
which is parametrically smaller than the mass of the Higgs boson.  
While it is not possible to use an
$x$-dependent  factorization scale without resorting to  
a full resummation  program, in the fixed order 
calculation we can attempt to do this on average. 
This choice {\it decreases} the NNLO corrections and  the Higgs boson 
production cross-section {\it increases} as compared 
to conventional choice of the scales, $\mu_r = \mu_f = m_H$.
\begin{figure}[htb]
\hspace*{1 mm}
\begin{minipage}{16.cm}
\begin{picture}(100,0)
\put (140,0) {$\sigma(p  p \to H + X)~[{\rm pb}],~~\sqrt{s} 
 = 14~{\rm TeV}$}
\put (100,-160) {$\mu,~~{\rm GeV}$}
\put (350,-160) {$\mu,~~{\rm GeV}$}
\end{picture}
\begin{tabular}{cc}
\psfig{figure=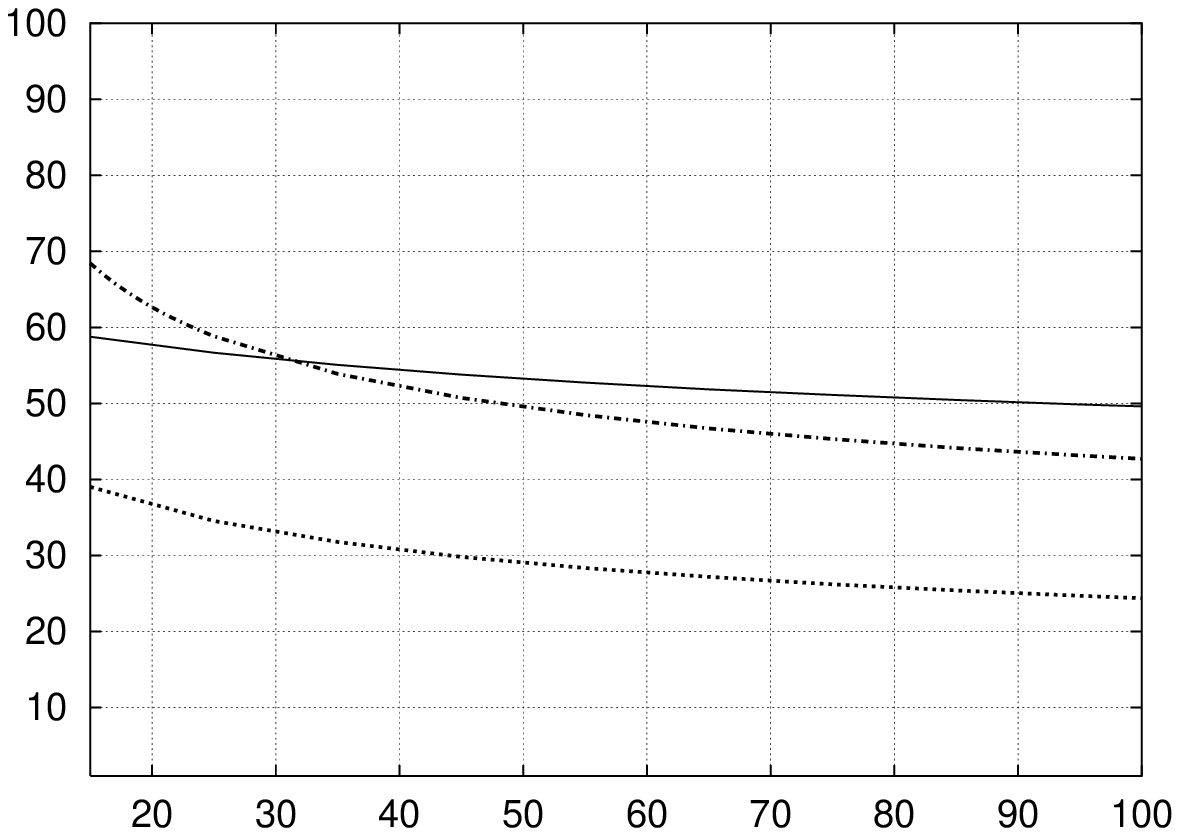,width=78mm}
&\hspace*{0mm}
\psfig{figure=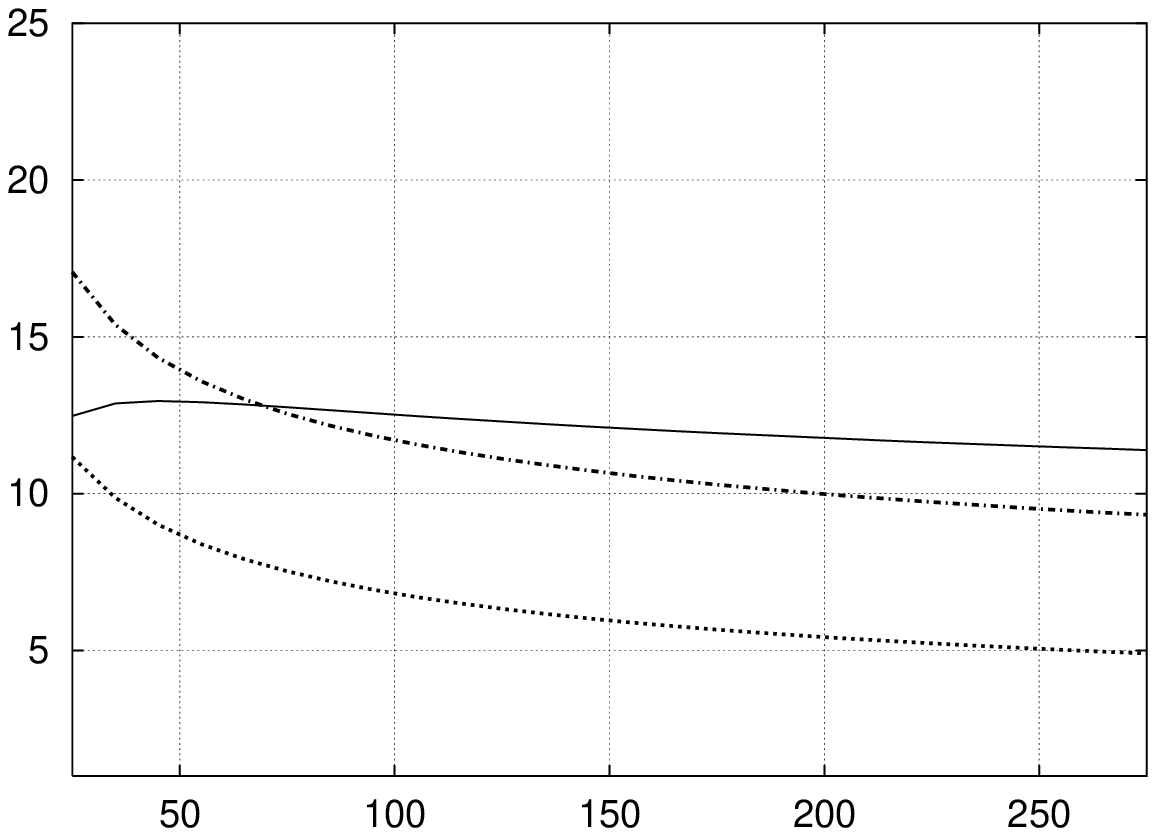,width=78mm}
\end{tabular}
\end{minipage}
\vspace*{0.5cm}
\caption{
The Higgs boson production cross-section at the LHC at 
leading (dotted), next-to-leading (dashed-dotted) 
and next-to-next-to-leading (solid)
order as the function of factorization and renormalization scale $\mu$.
The mass of the Higgs boson is $115~{\rm GeV}$ for the left  
and $275~{\rm GeV}$ for the right plot. }
\label{fig:plot2}
\end{figure}
We demonstrate this behavior with two  examples in Fig.~\ref{fig:plot2},
where we plot the production cross-sections for  $m_H = 115~{\rm GeV}$ and 
$m_H = 275~{\rm GeV}$. 
We equate the renormalization and factorization 
scales and vary the factorization scale from $\mu = 15~{\rm GeV}$
up to the mass of the Higgs boson. These plots illustrate that 
for smaller values of $\mu$, the NLO cross-section 
increases more rapidly than the NNLO cross-section, and the 
difference between the NLO and the NNLO results becomes smaller.
Therefore, the convergence of the perturbative series is improved 
for smaller values of the factorization scale. 

If we adopt this argument and restrict our analysis to small $\mu$ 
we find a Higgs production cross-section of
$55 \pm 5~{\rm pb}$ for $m_H = 115~{\rm GeV}$, 
a somewhat larger value than obtained with the conventional
scale choice $\mu=m_H$. It is interesting that recent
studies~\cite{catanithr}
of the threshold resummed cross-section for Higgs boson production, matched to 
the NNLO calculation, detect a similar increase  as compared to fixed order
calculations with $\mu=m_H$. 

\section{Conclusions}
\label{sec:conclusions}
In this paper, we have studied the Higgs boson production cross-section 
in hadron-hadron collisions. The main contribution to the hadronic
cross-section originates from gluon-gluon fusion, 
which we have computed at NNLO ($\O{\alpha_s^4}$) 
in perturbative QCD. The other partonic 
production channels, $qg \to H+X$, $q \bar q \to H+X$, $qq \to H+X$ and 
$qq' \to H+X$, were also studied to order $\alpha_s^4$. 

We have presented explicit analytic expressions for the partonic 
cross-sections valid within the heavy top-quark approximation. 
Finally, we have calculated  the cross-section for direct 
Higgs boson production at the Tevatron and the LHC 
by performing a numerical convolution 
of the partonic cross-sections with the MRST 2002 NNLO set~\cite{msrt} 
of parton distribution functions. The residual scale dependence of the  
NNLO cross-section is  approximately $\sim 15\%$ for  the LHC and 
$\sim 23\%$ for the Tevatron. 
The NNLO $K$-factors are fairly  large for both the LHC and the Tevatron. 
Nevertheless, the cross-section increases less from NLO to NNLO than from LO
to NLO, indicating a slow convergence of the perturbative expansion.  

While this calculation was in progress, Harlander and Kilgore~\cite{hkhard} 
obtained an approximation of the partonic cross-sections by expanding 
around the Higgs boson production threshold. Our results for the partonic
cross-sections  are in complete agreement with their expansion. 
Also, we find a very good agreement of our numerical results for the NNLO 
Higgs production cross section with the results reported in~\cite{hkhard}.

In Section~\ref{sec:numerics}, we have argued that it is more appropriate
to choose smaller values of the factorization scale $\mu$ than 
the conventional choice $\mu=m_H$. Then, the NNLO corrections {\it decrease}, 
indicating a better convergence of the perturbative series.  Moreover, with
this choice, the fixed order results are in better agreement with recent 
estimates of the cross-section  based on threshold  resummation~\cite{catanithr}.
  
We have suggested a method for the algorithmic evaluation of inclusive 
phase-space integrals.  This method combines the Cutkosky rules with  
integration-by-parts reduction algorithms to achieve a systematic reduction 
of phase-space integrals to a few master integrals. We have also shown how to 
compute these master integrals using differential equations produced with the 
IBP reduction algorithms. 
 
The techniques discussed in this paper can be used to  
compute higher order corrections to other inclusive processes 
of direct phenomenological interest. We are also confident that 
this approach can be generalized  to enable  the calculation  
of differential  distributions. In fact, a connection between phase-space  
integrals  with a modified measure, such as the integrals that  appear  
in the evaluation of  invariant mass, energy and angular distributions,  
and loop integrals with unconventional propagators exists. This connection 
can be used to automate the calculation of differential distributions 
following the lines of Section~\ref{sec:method}. The above ideas 
will be the subject of more detailed studies in future work. 

{\it \bf Acknowledgments}
We would like to thank Robert Harlander and Bill Kilgore for comparisons
with their results of Ref.~\cite{hkhard}. We are grateful to Frank Petriello
for careful reading of the manuscript and invaluable suggestions.  
This research was supported by the DOE under grant number
DE-AC03-76SF00515.
\newpage
\section*{Appendix}
In this Appendix, some  formulae used in our calculation are given.

\subsection{Splitting functions}
We first summarize the formulae for the space-like splitting functions.
\ba
&& P_{gg}^{(0)} = 
\left ( \frac{11}{4}-\frac{1}{6}n_f \right )\delta(1-x)
+3 \left ( x(1-x)+\frac{(1-x)}{x}-1 \right )
+3 \left [ \frac{1}{1-x}\right ]_+,
\nonumber \\
&& P^{(0)}_{gq} = \frac{2}{3}\frac{(1+(1-x)^2)}{x},
\nonumber \\
&& P^{(0)}_{qg} = \frac{1}{4} \left (x^2 + (1-x)^2 \right),
\nonumber \\
&& P^{(0)}_{qq} = \frac{2}{3} \left ( \frac{3}{2}\delta(1-x) 
-1 - x + 2\left [\frac{1}{1-x} \right ]_+
\right ),
\nonumber \\
&& P^{(1)}_{gg} = 
\left ( 6+ \frac{27}{4}\zeta_3 \right ) \delta (1-x)
+\left ( \frac{67}{4} - \frac{9}{2}\zeta_2 \right ) 
\left [ \frac{1}{1-x}\right ]_+
+ \frac{9 (x^2+x+1)^2}{x(1+x) }\Li_2(-x)
\nonumber \\
&& + \frac{9 (3+2x^2+4x+2 x^3)\zeta_2}{2(1+x)}
+ \frac{9 (x^2-x-1)^2}{2(1-x^2)}\ln^2 x
-\frac{25}{8}- \frac{109x}{8}
\nonumber \\
&& +\left ( \frac{9(x^2+x+1)^2}{x(1+x)}\ln(1+x)
- \frac{9 (x^2-x+1)^2}{x(1-x)}\ln(1-x)
- \frac{75}{4}+\frac{33}{4}x-33x^2
\right ) \ln(x)
\nonumber \\
&& + n_f \left [-\frac{2}{3} \delta(1-x)
-\frac{5}{6}\left [ \frac{1}{1-x}\right ]_+
-\frac{1+x}{3}\ln^2(x)
-\left (\frac{3}{2}+\frac{13x}{6} \right )\ln(x)
+ \frac{(-9 x-9 x^2+109x^3-61)}{36x}
\right ],
\nonumber \\
&& P^{(1)}_{gq} = \frac{9x+23x^2+9+44x^3}{9x} 
-\frac{4 n_f (4x^2+5-5x)}{27x} +\frac{2(2+2x+x^2)\lid{-x}}{x}
+4\zd
\nonumber \\
&& +\frac{5(2-2x+x^2) \log^2(1-x)}{9x}
+\log(x) \left( 
\frac{2(2+2x+x^2)\log(1+x)}{x} -\frac{100}{9} -\frac{31x}{9}
-\frac{8x^2}{3} \right) 
\nonumber \\
&& +\log(1-x) \left(
-\frac{2(2-2x+x^2)}{x}(\log(x)+\frac{n_f}{9}) +\frac{31x^2+42-42x}{9x} 
\right)+\frac{(14+11x)\log^2(x)}{9}.
\ea

\subsection{Master Integrals} 
Below we list the master integrals required in this calculation. 
We denote $\hat s=p_{12}^2=(p_1+p_2)^2$,  $z = \M / \hat s$ and for 
simplicity we set $\hat s=1$. 

\subsubsection{Double-Virtual}
The master integrals for the double-virtual corrections can be expressed
in terms of Gamma functions with the exception of the cross-triangle which
has been calculated in~\cite{xtriangle}. For completeness, we list their
$\epsilon$-expansion below: 
\vspace{-0.7cm} \hfill \\
\begin{eqnarray}
\xtri{27} &\equiv&  \measureVV \frac{\delta(p_{12}^2-\M)}{k^2 (k+p_{12})^2
  (l+p_{12})^2 (l+p_1)^2 (k-l-p_1)^2 (k-l)^2} \nonumber \\
&& \nonumber \\
&=& 
\delta \left( 1-z\right) 
\Gamma  \left(1+ \epsilon \right)  ^{2} 
{\rm Re}\left(e^{-2i\pi \epsilon} \right)
\left[ 
\frac{1 }{{\epsilon}^{4}}
-\,{\frac {7 \zeta_2}{{\epsilon}^{2}}}
-27\,{\frac {\zeta_3 }{\epsilon}}
-{\frac {19}{60}}\,{\pi }
^{4} + {\cal O}(\ep)
\right],  
\end{eqnarray}
\vspace{-1cm} \hfill \\
\begin{eqnarray}
\tri{27} &\equiv&  \measureVV \frac{\delta(p_{12}^2-\M)}{k^2 (k+p_{12})^2
  (l+p_{12})^2 (k-l)^2} \nonumber \\
&& \nonumber \\
&&\hspace{-2cm}=\delta \left( 1-z\right) 
\Gamma  \left(1+ \epsilon \right)  ^{2} 
{\rm Re}\left(e^{-2i\pi \epsilon} \right) \frac{1}{2}
\Bigg[ 
\frac{1}{\ep^2} + \frac{5}{\ep} + 19 
+ \left(65 -8\zeta_3 \right) \ep  \nonumber \\
&& \hspace{-2cm} 
+ \left( 211-40\zeta_3- \frac{2 \pi^4}{15}\right) \ep^2 + {\cal O}(\ep^3)
\Bigg],
\end{eqnarray}
\vspace{-1cm} \hfill \\
\begin{eqnarray}
\suns{27} &\equiv&  \measureVV \frac{\delta(p_{12}^2-\M)}{k^2 
(l+p_{12})^2 (k-l)^2} \nonumber \\
&& \nonumber \\
 &&\hspace{-3cm}=\delta \left( 1-z\right) 
\Gamma  \left(1+ \epsilon \right)  ^{2} 
{\rm Re}\left(e^{-2i\pi \epsilon} \right) \frac{1}{2}
\Bigg[
-\frac{1}{4\ep} -\frac{13}{8} 
+\left(
\frac{\zd}{2} -\frac{115}{16}
\right) \ep 
\nonumber \\  && \hspace{-3cm} 
+ \left( 
\frac{5\zt}{2} -\frac{865}{32} +\frac{13 \zd}{4}
\right) \ep^2
+\left( 
-\frac{5971}{64} + \frac{115 \zd}{8} + \frac{65 \zt}{4} 
+ \frac{11 \zd^2}{10}
\right) \ep^3
+{\cal O}(\ep^4)
\Bigg],
\end{eqnarray}
\vspace{-1cm} \hfill \\
\begin{eqnarray}
\glass{27} &=& 
 \delta(p_{12}^2-\M) 
\left[ 
\int \frac{d^dk}{i\pi^{\frac{d}{2}}} \frac{1}{k^2 (k+p_{12})^2}
\right]^2 \nonumber \\
&& \nonumber \\
&& \hspace{-3cm}=\delta \left( 1-z\right) 
\Gamma  \left(1+ \epsilon \right)  ^{2} 
{\rm Re}\left(e^{-2i\pi \epsilon} \right) 
\Bigg[
\frac{1}{\ep^2}+\frac{4}{\ep} + 12- 2\zd 
+ \left(32- 8\zd -4\zt \right) \ep +
\nonumber \\
&& \hspace{-3cm}
+\left(
80-16\zt -24\zd -\frac{4 \zd^2}{5}
\right) \ep^2 + {\cal O}(\ep^3) 
\Bigg],
\end{eqnarray}
\vspace{-1cm} \hfill \\
\begin{eqnarray}
\cglass{27} &=&  \delta(p_{12}^2-\M) 
\left| 
\int \frac{d^dk}{i\pi^{\frac{d}{2}}} \frac{1}{k^2 (k+p_{12})^2}
\right|^2
\nonumber \\
&& \nonumber \\
&& \hspace{-3cm}=\delta \left( 1-z\right) 
\Gamma  \left(1+ \epsilon \right)  ^{2} 
\Bigg[
\frac{1}{\ep^2}+\frac{4}{\ep} + 12- 2\zd 
+ \left(32- 8\zd -4\zt \right) \ep +
\nonumber \\
&& \hspace{-3cm}
+\left(
80-16\zt -24\zd -\frac{4 \zd^2}{5}
\right) \ep^2 + {\cal O}(\ep^3) 
\Bigg].
\end{eqnarray}


\subsubsection{Real-Virtual}
For the real-virtual contributions we find six master integrals:
\vspace{-1cm} \hfill \\
\begin{eqnarray}
&& \YE{27}  =  \measureVR \frac{\delta\left(k^2-\M\right) 
\delta\left((k+p_{12})^2\right) }
{l^2 (l+p_{12})^2}
\nonumber \\
&& \nonumber \\
&& \nonumber \\
&& = \PS \LOOP {\rm Re}\left( e^{-i\pi\epsilon}\right) s^{-2\epsilon} (1-z)^{1-2\epsilon},
\end{eqnarray}
\vspace{-0.5cm} \hfill \\
\begin{eqnarray}
&& \YA{27}  = \measureVR \frac{\delta\left(k^2-\M\right) 
\delta\left((k+p_{12})^2\right) }
{(k-l)^2 (l+p_1)^2}  
\nonumber \\
&& \nonumber \\
&& \nonumber \\
&& = \PS \LOOP s^{-2\epsilon} \frac{1-2\epsilon}{1-3\epsilon}
\frac{\Gamma(1-2\epsilon)^2}{\Gamma(1-\epsilon)\Gamma(1-3\epsilon)} (1-z)^{1-3\epsilon},
\end{eqnarray}
\vspace{-1cm} \hfill \\
\begin{eqnarray}
&& \YB{27}  =  \measureVR \frac{\delta\left(k^2-\M\right) 
\delta\left((k+p_{12})^2\right) }
{(k-l)^2 l^2}
\nonumber \\
&& \nonumber \\
&& \nonumber \\
&&= \PS \LOOP s^{-2\epsilon} {\rm Re} \left(e^{-i\pi\epsilon}\right) z^{-\epsilon} (1-z)^{1-2\epsilon},
\end{eqnarray}
\vspace{-0.5cm} \hfill \\
\begin{eqnarray}
&&\YC{27}  = \measureVR \frac{ \delta\left(k^2-\M\right) 
\delta\left((k+p_{12})^2\right) }
{(k-l)^2 (l+p_1)^2 l^2},
\nonumber \\
&& \nonumber \\
&& \nonumber \\
&& = \PS \Gamma(1+\ep) \frac{z^{-2\ep}}{1-2\ep} \Bigg\{ \frac{1}{\ep} \left[  
\lid{z} +\frac{\log^2(z)}{2} -\zd
\right]  + 5 \Sab{z} \nonumber \\
&&\hspace{0cm} - 4 \lit{z} + 4 \log(z) \lid{z} 
+\frac{\log^3(z)}{2} +2 \zd \log(z) -\zt
\nonumber \\
&& \hspace{0cm}
+\O{\ep} \Bigg\},  
\end{eqnarray}
\vspace{-1cm} \hfill \\
\begin{eqnarray}
&& \YD{27}  =  \measureVR \frac{\delta\left(k^2-\M\right) 
\delta\left((k+p_{12})^2\right) }
{(k+p_1)^2 (l+p_{12})^2 (l+p_2)^2 l^2 (k-l)^2},
\nonumber \\
&& \nonumber \\
&& \nonumber \\
&&= -\PS  \frac{\Gamma(1+\ep)}{3(1-2\ep)} \Bigg\{ 
(1-z)^{1-3\ep} \left[ 
\frac{1}{\ep^3} - \frac{2\zd}{\ep} -6\zt -11 \zq \ep + \O{\ep^2}
\right]
\nonumber \\
&&+ (1-z)^{1-2\ep} \left[ 
\frac{\log(z)}{\ep^2} 
+ \frac{\lid{z}-\log^2(z) -\zd}{3 \ep} \right. 
\nonumber \\ 
&& \left. + \frac{ 
\Sab{z} -2 \lit{z} - 8 \zd \log(z) + \zt}{3}
+\O{\ep}
\right] \Bigg\} ,
\end{eqnarray}
\vspace{-1cm} \hfill \\
\begin{eqnarray}
&& \YF{27} = \measureVR \frac{ \delta\left(k^2-\M\right) 
\delta\left((k+p_{12})^2\right) }
{(l+p_{12})^2 (l+p_1)^2 (k-l-p_1)^2 (k-l)^2}
\nonumber \\
&& \nonumber \\
&& \nonumber \\
 &&= -2 \PS \frac{\Gamma(1+\ep)}{1-2\ep} (1-z)^{-1-4\ep} 
\Bigg\{ \left[ 
\frac{1}{\ep^3} -\frac{6 \zd}{\ep} - 16 \zt -18\zq \ep + \O{\ep^2} 
\right].
\nonumber \\
&&\hspace{0cm} + \left[
\frac{\log(z)}{\ep^2} -\frac{\lid{z}-\zd}{\ep} 
+ \Sab{z}-2 \lit{z} -\frac{\log^3(z)}{3} +\zt  +\O{\ep}
\right] \Bigg\},
\end{eqnarray}
where the common factors $\PS$ and $\LOOP$ are:
\be
\PS = \frac{\pi^{\frac{3}{2}-\ep}}{2^{3-2\ep} \Gamma(\frac{3}{2}-\ep)}
\ee
and 
\be
 \LOOP = \frac{\Gamma(\ep) \Gamma(1-\ep)^2}{\Gamma(2-2\ep)}.
\ee

\subsubsection{Double-Real}
We find the following master integrals for the double-real contributions:
\begin{eqnarray}
&& \Xone{27} \equiv  
\measureRR \delta((k-l)^2-\M) \delta((l+p_1+p_2)^2) \delta(k^2) 
\nonumber \\
&& \nonumber \\
&& \nonumber \\
&&\hspace{-0.5cm} = \frac{\PS^2 \Gamma(1-2\ep)^2}{\Gamma(1-4\ep)} \Bigg\{ 
\frac{1-2\ep}{2 (3-4\ep) (1-4\ep)} (1-z)^{3-4\ep} 
+ (1-z)^{-4\ep} \Bigg[-\frac{1}{6}(1-z)\left( z^2-5z-2\right) +
z\log(z) 
\nonumber \\
&&\hspace{-0.5cm} -\ep 
\left(\frac{5}{36}(1-z)(4z^2-17z-5)+\frac{z (z+2)}{2}\log(z) 
+\frac{z\log(z)^2}{2}+ 4z \lid{1-z}
\right) \nonumber \\
&&\hspace{-0.5cm} + \ep^2 \left(
8z \Sab{1-z}- 16 z \lit{1-z}+\frac{z\log(z)^3}{6}+\frac{56z^3}{27}
+\frac{z(2+z)}{4}\log(z)^2 \right. \nonumber \\
&& \hspace{-0.5cm} 
\left.
-\frac{691}{72}z^2+\frac{56z}{9}+\left(z^2+4z+1\right) \lid{1-z} 
+ \left( 
6 \lid{1-z} -\frac{5z+8}{4}
\right) z\log(z) \right. \nonumber \\
&& \hspace{-0.5cm} \left.
+\frac{281}{216} \right) + \O{\ep^3} \Bigg] \Bigg\},
\end{eqnarray}
\vspace{-1cm} 
\begin{eqnarray}
&& \Xfourteen{27} \equiv  
\measureRR \delta((k-l)^2-\M) \delta((l+p_1+p_2)^2) \delta(k^2) l^2 
\nonumber \\
&&\nonumber \\
&& \nonumber \\
&& \hspace{0cm} = \frac{\PS^2 \Gamma(1-2\ep)^2}{\Gamma(1-4\ep)} (1-z)^{-4\ep} 
\Bigg\{ (1-z)^3 
+ \ep \left[ 
(3-z)z^2\log(z) \right. \nonumber \\
&& \hspace{0cm} \left.
+\frac{1}{2}(1-z)(5z^2-6z+5)
\right] 
+ \ep^2 \left[ 
\frac{1}{2}(z-3)z^2\log(z)^2 -\frac{1}{2} (5z^2-7z+4)z \log(z)
\right. \nonumber \\ 
&& \hspace{0cm} \left. +2 (z+1)(z^2-4z+1) \lid{1-z} 
+\frac{1}{4}(1-z)(27z^2-26z+27) \right]
\nonumber \\ 
&&\hspace{0cm} 
+\ep^3 \left[ 
-\frac{1}{6}(z-3)z^2\log^3(z)-6(z-3)z^2\lid{1-z}\log(z) 
+\frac{1}{4} z(5z^2-7z+4)\log^2(z)
\right. \nonumber \\
&& \hspace{0cm} \left. 
-\frac{1}{4}(27z^2-25z+28)z \log(z)+ (1+z)(5z^2-8z+5)\lid{1-z}
\right. \nonumber \\
&& \hspace{0cm}
\left. + 8 (1+z) (z^2-4z+1)\lit{1-z} + (-10z^3+30z^2-6z+2) \Sab{1-z}
\right. \nonumber \\
&& \hspace{0cm}
\left. +\frac{1}{8}(1-z)(153z^2-142z+153) \right] + \O{\ep^4}
\Bigg\},
\end{eqnarray}
\vspace{-1cm} 
\begin{eqnarray}
&& \Xten{27} \equiv  
\measureRR 
\frac{\delta((k-l)^2-\M) \delta((l+p_1+p_2)^2) \delta(k^2)}
{l^2 (k+p_1+p_2)^2}
\nonumber \\
&& \nonumber \\
&& \nonumber \\
&& =\PS^2 z^{-\ep}(1+z)^{-\ep} \Bigg\{ 
-2\lid{-z}-\zd +\frac{\log(z)}{2}\left(\log(z)-4\log(1+z)\right) \nonumber \\
&& \hspace{0cm} + \ep \left[
-8\log(1+z)\lid{z} + \frac{\log(z)}{6} \left(
\log^2(z)-6 \log(z)\log(1+z) -6 \log(1+z)^2 
\right. \right. \nonumber \\ 
&& \hspace{0cm}
\left. \left. -6 \log(z)+ 24 \log(1+z) \right) 
+4 \lit{z}+(4-3\log(z)) \lid{-z}+10 \Sab{-z}
\right. \nonumber \\
&& \hspace{0cm}
\left. 
-4 \Sab{z^2} +8 \Sab{z} + 3 \lit{-z} 
+ \left( 2+8 \log(1+z)-4\log(z) \right)\zd -7\zt \right] 
\nonumber \\
&& \hspace{0cm} + \O{\ep^2}\Bigg\},
\end{eqnarray}
\vspace{-1cm} 
\begin{eqnarray}
&& \Xthirteen{27} \equiv  \measureRR 
\frac{\delta((k-l)^2-\M) \delta((l+p_1+p_2)^2) \delta(k^2)}
{l^2 (l+p_2)^2 (k+p_1)^2 (k+p_1+p_2)^2}
 \nonumber \\
&& \nonumber \\
&& \nonumber \\
&&= \PS^2 z^{-1-2\ep} \Bigg\{ 
-\frac{\log(z)}{\ep^2} + \frac{1}{\ep} \left[ 
2\zd -2 \lid{z^2} -4\log(z)\left(\log(1+z)-1\right)
\right]
\nonumber \\
&& \hspace{0cm}
-\log(z)\left( 7 \log(z) \log(1+z) -2 \log(1+z)^2 -16 \log(1+z)+ 4\right)
+ 20 \lit{z}
\nonumber \\
&& \hspace{0cm}
+(16-10\log(z)-16\log(1+z))\lid{z} +  (16-14\log(z)+4\log(1+z)) \lid{-z}
\nonumber \\
&& \hspace{0cm}
+20 \Sab{-z} -8 \Sab{z^2} + 14 \lit{-z}
+ \zd \left(-6\log(z)+18\log(1+z)-8 \right)
-4 \zt 
\nonumber \\
&& \hspace{0cm}
+ \O{\ep} \Bigg\},
\end{eqnarray}
\vspace{-1cm} 
\begin{eqnarray}
&& \Xeleven{27} \equiv  \measureRR 
\frac{\delta((k-l)^2) \delta((l+p_1+p_2)^2) \delta(k^2-\M)}
{l^2 (l+p_1)^2 (k+p_1)^2 (k+p_1+p_2)^2}
\nonumber \\
&& \nonumber \\
&& \nonumber \\
&& =\PS^2 (1+z)^{-1-2\ep} \Bigg\{ 
-\frac{2 \log(z)}{\ep^2} +\frac{1}{\ep} \left[
10\zd +8 \log(z)-\log(z)^2 \right. \nonumber \\
&& \hspace{0cm}
\left. + 4 \lid{-z} -8 \lid{z} \right]+\frac{4}{3}\log^3(z)+
4 \log^2(z) +34 \zt -12 \lit{-z} -24 \Sab{-z}
\nonumber \\
&& \hspace{0cm} + \left(8\lid{-z}-8-4\lid{z}+10\zd \right) \log(z)
+8 \Sab{z^2}-48 \Sab{z} -40 \zd
\nonumber \\ 
&& \hspace{0cm}
 -16 \lid{-z}+ 32 \lid{z} + \O{\ep} \Bigg\},
\end{eqnarray}
\vspace{-1cm} 
\begin{eqnarray}
&& \Xtwentyfive{27} \equiv  \measureRR 
 \frac{\delta((k-l)^2) \delta((l+p_1+p_2)^2) \delta(k^2-\M)}
{l^2 (l+p_2)^2 (k+p_1)^2 (k+p_1+p_2)^2} \nonumber \\
&& \nonumber \\
&& \nonumber \\
&& = \PS^2 (1-z)^{-1-4\ep} \Bigg\{ 
-\frac{1}{\ep^3} + \frac{4}{\ep^2} + \frac{4-4\zd}{\ep}
+ 16 \left(\zt-\zd \right) + 4\ep [ 4\zd -16\zt \nonumber \\
&& \hspace{0cm}
+9 \zq ] + \O{\ep^2} \Bigg\} + \PS^2 (1-z)^{-1-2\ep} \Bigg\{
\frac{-2\log(z)}{\ep^2} + \frac{8 \log(z) -4 \lid{z}+ 4 \zd}{\ep}
+ 6 \lit{z} \nonumber \\
&& \hspace{0cm}
-10 \Sab{z}+ 4 \zt + \left(16-2\log(1-z)-4\log(z)\right) \lid{z}
+ \zd \left( 6\log(z)+2 \log(1-z)-16\right) \nonumber \\
&& \hspace{0cm} + \O{\ep} \Bigg\},
\end{eqnarray}
\vspace{-1cm} 
\begin{eqnarray}
&& \Xfive{27} \equiv  \measureRR 
\frac{\delta((k-l)^2) \delta((l+p_1+p_2)^2) \delta(k^2-\M)}
{(l+p_2)^2 (l+p_1)^2 (k+p_1)^2 (k+p_2)^2}
\nonumber \\
&& \nonumber \\
&& \nonumber \\
&& = \PS^2 (1-z)^{-1-4\ep} \Bigg\{ 
\frac{-1}{\ep^3} +  \frac{4}{\ep^2} 
+ \frac{4\zd -\frac{10}{3}-\frac{2}{3}z}{\ep}
+ \frac{20}{9}(1-z)+ 16 \zt \nonumber \\
&& \hspace{0cm} -16 \zd + \ep \left[ 
\frac{224}{27}(1-z) + \frac{8}{3}(5+z) + 36 \zq -64 \zt
\right] + \O{\ep^2} \Bigg\} \nonumber \\
&& \hspace{0cm} +\PS^2 (1-z)^{-1-2\ep}
\Bigg\{ \frac{-2 \log(z)}{\ep^2}
+\frac{8\log(z)- 4 \lid{z} +4 \zd-\frac{2}{3}(1-z)}{\ep}
+ 4 \zt - \frac{20}{9}(1-z) \nonumber \\
&& \hspace{0cm} + 6 \lit{z} + 16 \lid{z} -10 \Sab{z}
+ \frac{\log^3(z)}{3}- \log^2(z)\log(1-z) -16 \zd \nonumber \\
&& \hspace{0cm}+\log(z) \left[ 
6\zd - log(1-z)^2 - 8 - 4 \lid{z}
\right] + \log(1-z) \left[ \frac{4(1-z)}{3} +2 \zd-2 \lid{z}\right]
\nonumber \\
&& \hspace{0cm} + \O{\ep} \Bigg\},
\end{eqnarray}
\vspace{-1cm} 
\begin{eqnarray}
&& \Xeight{27} \equiv  \measureRR 
\frac{\delta((k-l)^2-\M) \delta((l+p_1+p_2)^2) \delta(k^2)}
{(l+p_2)^2 (l+p_1)^2 (k+p_1)^2 (k+p_2)^2}
\nonumber \\
&& \nonumber \\
&& \nonumber \\
&& \frac{\Gamma(1-2\ep)^2 \PS^2}{\Gamma(1-4\ep)} (1-z)^{-1-4\ep} 
\Bigg\{ -\frac{4}{\ep^3} + \frac{16}{\ep^2} + \frac{z(z-8)-89}{6\ep}
\nonumber \\
&& \hspace{0cm}
+ \frac{1}{9}(z-1)(1+2z)+ \frac{2\ep}{27} (z-1)(13z-16) + \O{\ep^2} \Bigg\}
\nonumber \\
&& \hspace{0cm}
+ \frac{\Gamma(1-2\ep)^2 \PS^2}{\Gamma(1-4\ep)} (1-z)^{-1-2\ep} \Bigg\{
\frac{-2\log(z)}{\ep^2} + \frac{1}{\ep} \left[ 2\log(z)^2
+ \log(z)(4 \log(1-z)+8) \right. \nonumber \\ 
&& \hspace{0cm}
\left. +\frac{1}{6}(1-z)(z-7) \right] -\frac{4}{3} \log^3(z) -\log^2(z)
(8+2 \log(1-z))  +36 \left( \lit{z}-\zt\right) 
\nonumber \\ 
&& \hspace{0cm}
+ \frac{1}{9}(z-2z^2+1) -\log(z) \left( 
20\zd + 16 \lid{z} + 16 \log(1-z)-4 \log(1-z)^2-8
\right) \nonumber \\
&& \hspace{0cm}
+ \frac{1}{3}(z-1)(z-7)\log(1-z) + \O{\ep} \Bigg\},
\end{eqnarray}
\vspace{-1cm} 
\begin{eqnarray}
&& \Xthree{27} \equiv  \measureRR 
\frac{\delta((k-l)^2-\M) \delta((l+p_1+p_2)^2) \delta(k^2)}
{(k+p_1)^2 (k+p_2)^2}
\nonumber \\
&& \nonumber \\
&& \nonumber \\
&& = \PS^2 z^{-\ep}(1-z)^{-1-2\ep} \Bigg\{ \lid{1-z}+ \frac{\log^2(z)}{2} + 
\ep \left[ 6\lit{1-z} \right.  \nonumber \\
&& \hspace{0cm} 
\left. -7 \Sab{1-z} -2 \lid{1-z} \left(1+\log(1-z)+2\log(z)\right)
+ \frac{1}{6} \log^3(z) 
\right. \nonumber \\
&& \left. \hspace{0cm} -\log^2(z) (1+\log(1-z)) \right] + \O{\ep^2}  
\Bigg\},
\end{eqnarray}
\vspace{-1cm} 
\begin{eqnarray}
&& \Xsix{27} \equiv  \measureRR 
\frac{\delta(k^2-\M) \delta((l-p_1)^2) \delta((k+l+p_2)^2)}
{(k+p_1)^2(k+p_2)^2(k+l)^2l^2}
\nonumber \\
&& \nonumber \\
&& \nonumber \\
&& \nonumber \\
&& = \frac{\Gamma(1-2\ep)^2\PS^2}{\Gamma(1-4\ep)}(1-z)^{-1-4\ep}z^{-1-2\ep}
\Bigg\{ \Bigg[ -\frac{1}{\ep^3} + \frac{11+z(z-4)}{2\ep^2} \nonumber \\
&& \hspace{0cm} 
+ \frac{82z-21z^2-85}{6\ep} -\frac{211}{9}z+\frac{263}{18} 
+\frac{53}{6}z^2 + \ep \left[ 
\frac{587}{27}z-\frac{811}{54} -\frac{121}{18}z^2
\right] + \O{\ep^2} \Bigg] \nonumber \\
&& \hspace{0cm} 
+\Bigg[ -\frac{\log(z)+\frac{1}{2}(z-1)(z-3)}{\ep^2}
+\frac{1}{\ep} \left[ 
\frac{1}{6}(z-1)(21z-61)+4\log(z)-4\lid{1-z}
\right] \nonumber \\
&& \hspace{0cm} 
-4\Sab{1-z}-10\lit{1-z}+ (16-10\log(z))\lid{1-z}-4\log(z)
\nonumber \\
&& \hspace{0cm}
+\frac{1}{18}(1-z)(159z-263) + \O{\ep} \Bigg] \Bigg\},
\end{eqnarray}
\vspace{-1cm} 
\begin{eqnarray}
&& \Xthirty{27} \equiv  \measureRR 
\frac{\delta(k^2-\M) \delta((l-p_2)^2) \delta((k+l+p_1)^2)}
{(k+p_1+p_2)^2(k+l)^2 l^2}
\nonumber \\
&& \nonumber \\
&& \nonumber \\
&& \nonumber \\
&& =(1-z)^{-1-4\ep}(1-2\ep)^2 \PS^2\Bigg\{ 
\left[-\frac{3}{\ep^3}+\frac{14\zd}{\ep}+58\zt +119 \zq \ep +\O{\ep}
\right] \nonumber \\
&& \hspace{0cm} + z^{-\ep} \Bigg[
-\frac{\log(z)}{\ep^2} +20 \left( \lit{z}-\zt \right) 
-6\zd \log(z) -10 \log(z) \lid{z} -\frac{\log^3(z)}{6} 
\nonumber \\
&& \hspace{0cm} +\O{\ep}
\Bigg] \Bigg\}, 
\end{eqnarray}
\vspace{-1cm} 
\begin{eqnarray}
&& \Xtwentyone{27} \equiv  \measureRR 
\frac{\delta(k^2-\M) \delta((k+l+p_1+p_2)^2) \delta(l^2)}
{(l+p_1)^2(k+p_2)^2(k+l+p_2)^2(l+p_2)^2}
\nonumber \\
&& \nonumber \\
&& \nonumber \\
&& \nonumber \\
&& =\PS^2 (1-2\ep)^2 z^{-2\ep} \Bigg\{
-\frac{\log(z)}{\ep^2} +\frac{4\zd-\lid{z} -\log^2(z)}{\ep}
\nonumber \\
&& \hspace{0cm} +12\lit{z}-16\Sab{z}+4\zt +2\zd \log(z) -10 \log(z)
\lid{z} -\frac{2}{3}\log^3(z)
\nonumber \\
&& \hspace{0cm} +\O{\ep} \Bigg\},
\end{eqnarray}
\vspace{-1cm} 
\begin{eqnarray}
&& \Xeighteen{27} \equiv  \measureRR 
\frac{\delta(k^2) \delta((l-p_2)^2) \delta((k+l+p_1)^2 -\M)}
{(k+p_1+p_2)^2(k+l)^2 l^2}
\nonumber \\
&& \nonumber \\
&& \nonumber \\
&& = \PS^2 (1+z)^{-1-4\ep}z^\ep (1-2\ep)^2 \Bigg\{
-\frac{\log(z)}{\ep^2} + \frac{6\zd +4\lid{-z}-4\lid{z}}{\ep}
\nonumber \\ 
&& \hspace{0cm} 
+ 2\log(z)\lid{z^2}  + 6\zd \log(z) +\frac{3\log^3(z)}{2}
-8\lit{z}  -32 \Sab{-z} +8 \Sab{z^2}
 \nonumber \\
&& \hspace{0cm} -32\Sab{z} -8\lit{-z} +30 \zt +\O{\ep} \Bigg\},
\end{eqnarray}
\vspace{-1cm} 
\begin{eqnarray}
&& \Xnine{27} \equiv  \measureRR 
\frac{\delta(k^2) \delta((l+p_1+p_2)^2) \delta((l-k)^2-\M)}
{(l+p_1+p_2)^2(k+p_1+p_2)^2(l+p_1)^2}
\nonumber \\
&& \nonumber \\
&& \nonumber \\
&& \nonumber \\
&& = \PS^2 z^{-2\ep} \left[ 
\frac{\log(z)^2}{2\ep} - 4 \zt -4\zd \log(z) + 4\lit{z} +\frac{\log^3(z)}{2}
+\O{\ep}
\right],
\end{eqnarray}
\vspace{-1cm} 
\begin{eqnarray}
&& \Xsixteen{27} \equiv  \measureRR 
\frac{\delta(k^2) \delta((l+p_1+p_2)^2- \M) \delta((l-k)^2)}
{(l+p_1+p_2)^2(k+p_1+p_2)^2(l+p_1)^2}
\nonumber \\
&& \nonumber \\
&& \nonumber \\
&& \nonumber \\
&& =\PS^2 z^{-2\ep} \left[
2\zt + \zd \log(z) + \frac{\log^3(z)}{6} + \log(z) \lid{z} -2 \lit{z}
+\O{\ep}
\right],
\end{eqnarray}
\vspace{-1cm} 
\begin{eqnarray}
&& \Xseven{27} \equiv  \measureRR 
\frac{\delta(k^2) \delta((l+p_1+p_2)^2) \delta((l-k)^2-\M)}
{(l+p_1)^2(k+p_1)^2}
\nonumber \\
&& \nonumber \\
&& \nonumber \\
&& \nonumber \\
&& =\PS^2 z^{-\ep}(1-2\ep)^2 \Bigg[
-\frac{\log(z)}{\ep^2} + \frac{4\zd -4\lid{z}}{\ep}
+12 \lit{z} -16 \Sab{z} 
\nonumber \\
&& 
-6 \log(z) \lid{z} - 2 \zd \log(z) - 
\frac{\log^3(z)}{6} + 4\zt + \O{\ep}
\Bigg],
\end{eqnarray}
\begin{eqnarray}
&& \Xnineteen{27} \equiv  \measureRR 
\frac{\delta(l^2) \delta((k+p_1+p_2)^2) \delta((l-k)^2-\M) (l+p_1)^2}
{(l+p_1+p_2)^2(k+p_1)^2}
\nonumber \\
&& \nonumber \\
&& \nonumber \\
&& \nonumber \\
&& = \PS^2 \Bigg\{ -\frac{\log^3(z)}{6} -(1-z)\left[1+\frac{\log^2(z)}{2} +\lid{1-z}\right] 
-\log(z) \Big[1+ \lid{1-z} \Big] 
\nonumber \\
&& -2\Sab{1-z} + \O{\ep}
\Bigg\},
\end{eqnarray}
\vspace{-1cm} 
\begin{eqnarray}
&& \Xtwenty{27} \equiv  \measureRR 
\frac{\delta((k+l+p_1+p_2)^2-\M) \delta(k^2) \delta(l^2)}
{(l+p_1)^2(l+p_1+p_2)^2(k+l+p_2)^2(k+p_2)^2}
\nonumber \\
&& \nonumber \\
&& \nonumber \\
&& \hspace{0cm} = \frac{\Gamma(1-2\ep)^2 \PS^2}{\Gamma(1-4\ep)} z^{-1-2\ep} (1-2\ep)^2 \Bigg\{
\left[ \frac{1}{2\ep^3} + \left( 
\frac{36\zd^2}{5} + 8 \zd -32\zt
\right) \ep + \O{\ep^2}\right] (1-z)^{-4\ep}
\nonumber \\ 
&& \hspace{0cm}
+\frac{(1+2z) \log(z)}{(1+z) \ep^2} 
+\frac{1}{\ep} 
\Bigg[
\frac{3z\log^2(z)}{1+z} 
+\frac{(1-5z)\zd}{1+z} 
+\frac{4z\lid{z}}{1+z}
+\frac{2(1-z)\lid{-z}}{1+z}
\nonumber \\ 
&& \hspace{0cm}
+\log(z) \left( 
-\frac{4(1+2z)}{1+z}
-4\log(1-z)
+\frac{2(1-z)\log(1+z)}{1+z}
\right)
\Bigg]
+\frac{1}{1+z} \Bigg[
(26z-6) \Sab{z} 
\nonumber \\
&& \hspace{0cm}
+ (16z-16) \Sab{-z}
+4(1-z)\Sab{z^2} -(6z+14)\lit{z} -4(z+3)\lit{-z}
+(9-21z)\zt \nonumber \\
&& \hspace{0cm}
+\left[ 
10(1-z)\log(z)-12(1-z)\log(1+z) -2(1+z)\log(1-z)
\right] \zd
\nonumber \\
&& \hspace{0cm}
+\left[ 
(12-4z) \log(z) +8(z-1) \log(1+z)
\right] \lid{-z}+3z\log^3(z)
\nonumber \\
&& \hspace{0cm}
+ \left[
4(1+3z)\log(z) +8(1-z)\log(1+z) +2(1+z)\log(1-z)
\right] \lid{z}
\nonumber \\
&& \hspace{0cm}
+\left[
6(1-z)\log(1+z) -3(1+z)\log(1-z) 
\right] \log^2(z) + \left[
4(z-1)\log^2(1+z) \right. \nonumber \\
&& \hspace{0cm} \left. 
+9(1+z)\log^2(1-z) -16(1+z)\log(1-z) 
\right] \log(z)
\Bigg]
\Bigg\}. 
\end{eqnarray}




\end{document}